\documentclass[a4paper,11pt]{article}
\usepackage{jinstpub} 
\usepackage{multirow}

\title{Experimental characterisation of a combined LVDT position sensor and voice-coil actuator for gravitational wave detectors}

\author[1]{K. A. Kukkadapu\note{Corresponding author.},}
\author{P. Li,}
\author{H. Van Haevermaet,}
\author{A. N. Koushik,}
\author{W. Beaumont}
\author{and N.~van Remortel}
\affiliation{University of Antwerp,\\Groenenborgerlaan 171, 2020 Antwerpen, Belgium}

\emailAdd{KumarAkhil.Kukkadapu@uantwerpen.be}

\abstract{A detailed characterisation of a combined Linear Variable Differential Transformer (LVDT) position sensor and voice-coil (VC) actuator designed for seismic isolation systems in gravitational wave detectors is presented. A dedicated experimental setup and a finite-element simulation framework were developed to measure and model a representative Einstein Telescope pathfinder Type-A LVDT+VC assembly. The setup employs a precision translation stage and balance to quantify LVDT displacement response and VC force output under controlled conditions. We found a good agreement between experiment and simulation: the measured LVDT response was determined with an uncertainty of 0.5\% and differed by only 1.3\% from the model prediction, demonstrating high linearity over a $\pm$5~mm range. In addition, the VC force measurements agreed within the total uncertainty: the maximum normalised force was determined with a precision of 2.3\% and matched the simulated value with only 0.6\% discrepancy. These results validate the combined sensor-actuator design and our measurement methodology. The demonstrated linear response and stable actuation confirm that this LVDT+VC device can be used for low-frequency suspension control. Our framework therefore provides a validated tool to optimise existing sensor and actuator designs, and to study novel prototypes for next-generation gravitational wave detectors.}


\arxivnumber{2603.06209} 

\begin{document}
\maketitle
\flushbottom

\section{Introduction}
\label{sec:introduction}

Ground-based gravitational wave (GW) observatories use complex seismic isolation systems~\cite{STOCHINO2009737,Acernese_2014,Heijningen:2019jmd, 10.1063/1.4866659,Akutsu:2021auw,Kirchhoff:2020vhb,Punturo:2010zz, VirgoSuperattenuator, SteeringFilter} to achieve the required detector sensitivity. Both position sensors and actuators play an important role in such attenuation chains to provide low-frequency active damping of suspension modes, and reduce the residual motion of the suspended mirrors or optical benches. They allow to precisely monitor and control slow drifts and maintain the alignment of different suspension stages.
For this purpose special Linear Variable Differential Transformers (LVDT) sensors~\cite{LowNoiseAccelerometer, tariq2002linear} and Voice Coil (VC) actuators~\cite{WANG2002563} were developed. These LVDTs are highly sensitive and linear sensors that obtain a relative position measurement via mutual induction. Due to their non-contact working principle, low-noise, nanometre resolution, and ultra-high vacuum (UHV) compatibility, they are well suited for applications in GW instrumentation.

A key distinction between the LVDTs used in industry and GW detectors is the adoption of a moving primary coil instead of a moving ferromagnetic core. In addition, the radius of the primary is significantly smaller than the radius of the secondary coils to allow for a transverse motion of the suspended mass. The movement of the excited primary (or inner) coil induces an electromotive force (emf) in the Maxwell pair arranged secondary (or outer) coils, which is detected as a differential voltage output. This contact-free operation makes them ideal for environments demanding minimal mechanical interference, while the linear response ensures accurate position measurements~\cite{nyce:2016}. When a permanent magnet is placed inside the primary coil, and the secondary coils are driven with a DC (or low-frequency) current, a VC actuator functionality can be added due to the generated Lorentz force. This creates a combined LVDT+VC system that enables precise position sensing and stable actuation with one device.

A detailed understanding and further optimisation of such position sensors and actuators in the suspension control system is essential to maximise the sensitivity and operational stability of next-generation gravitational wave detectors. This is especially important for the Einstein Telescope (ET)~\cite{Punturo:2010zz}, a future third-generation European GW observatory, which aims to extend the sensitivity down to lower frequencies (3~Hz) by several orders of magnitude~\cite{et_design_report_update2020}. In this regime, the noise, linearity, dynamic range, and stability of local position sensors and actuators directly limit the achievable seismic isolation performance~\cite{losurdoActiveControlsInterferometric2000, maggioreAngularControlNoise2025}, as sensor noise and actuator instabilities propagate through the feedback control loops and couple back into the suspension dynamics. This control noise can therefore affect low-frequency sensitivity and reduce robustness against environmental disturbances. In addition, a further optimisation of existing designs, or the development of novel sensor and actuator concepts, might simplify and improve construction methods, remove dependencies on environmental effects, or improve signal readout. 

This paper presents a new dedicated experimental setup and simulation framework to fully characterise the performance of LVDT sensors and VC actuators. The primary objective is to establish a validated, precise methodology that enables a direct comparison between experimental measurements and finite-element simulations. Beyond measurements of existing LVDT+VC designs, the setup allows the development, optimisation, and validation of novel sensor and actuator prototypes for future GW detectors. The experimental setup provides a controlled environment in which LVDT response, linearity, VC actuation strength, and stability can be accurately measured and calibrated. When combined with simulations, the framework can optimise LVDT+VC designs tailored to the requirements of specific suspension stages, e.g. by maximising sensor response, actuation efficiency, or dynamic range, while minimising noise contributions. In addition, the setup allows dedicated studies of environmental and operational effects, including stray magnetic fields, temperature variations, sensor cross-coupling, and electromagnetic interference between closely spaced devices. It also provides a platform to study improvements in analog amplification and digital signal processing, with the aim of enhancing the signal-to-noise ratio and readout stability.

As a representative case study, we use a reference LVDT+VC geometry currently deployed in ETpathfinder~\cite{ETpathfinderTDR, Utina:2022qqb}. ETpathfinder is a R\&D infrastructure designed to test advanced GW detector technologies in a low-noise, full-interferometer environment, specifically for the Einstein Telescope. The specific system investigated in this work is the ETpathfinder Type-A LVDT+VC combination, which is used to sense and control the horizontal degrees of freedom of the inverted pendulum stages and the rotational degrees of freedom of suspended optical benches. With our measurement setup we aim to precisely determine i) the LVDT response and linearity, ii) the VC actuation force and stability, and (iii) compare all results with predictions from finite-element simulations.  

This paper is structured as follows: section~\ref{sec:working_principle} briefly describes the working principle of a combined LVDT+VC system used in GW detectors, provides the ETpathfinder Type-A design parameters, and introduces the simulation framework that is used to obtain model predictions. Section~\ref{sec:exp-setup} presents the experimental measurement setup, and section~\ref{sec:LVDT} then shows the results of LVDT measurements of the Type-A assembly. The systematic uncertainties and a data driven correction factor are described in section~\ref{sec:systematics}, while measurement results of the VC actuator are presented in section~\ref{sec:VC_measurement}, followed by section~\ref{sec:conclusions} with the conclusions of our experimental characterisation. Throughout this paper, the terms ‘primary coil’ and ‘inner coil’ are used interchangeably, as are ‘secondary coil’ and ‘outer coil’.

\section{Combined LVDT and VC actuator working principle} 
\label{sec:working_principle}

\begin{figure}[ht]
    \centering
    \includegraphics[width=0.7\linewidth]{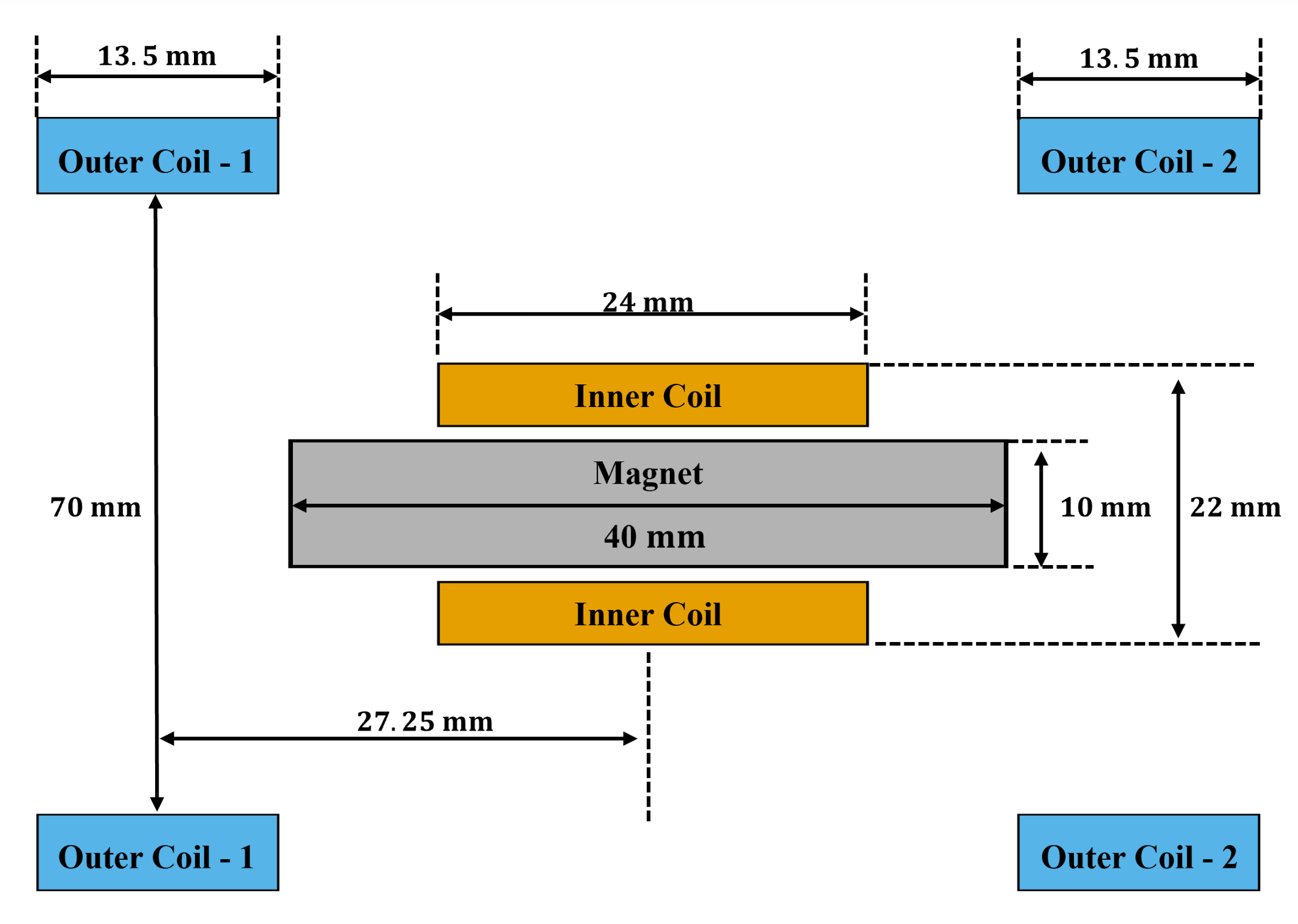}
    \caption{A schematic of a combined LVDT+VC system with all parts and their dimensions indicated.}
    \label{fig:lvdt geometry}
\end{figure}

The LVDT sensors used in GW detectors consist of three coaxially arranged helical coils, as indicated in figure~\ref{fig:lvdt geometry}. The central coil, commonly referred to as the primary (or inner) coil, is mechanically coupled to the object whose relative displacement is to be measured. The two outer coils, referred to as the secondary coils, are fixed to a stable reference structure and share the same central axis as the primary coil. They have a larger radius, are wound in opposite direction, and are connected in series. In the ideal case, the secondary coils are arranged in a Maxwell pair configuration satisfying $\rm D_{\rm s} = \sqrt{3}R_{\rm s}$~\cite{tariq2002linear}, where $\rm D_s$ is the distance between the secondary coils and $\rm R_s$ their radius.

When the primary coil is driven with a high frequency sinusoidal signal a time-varying magnetic field is produced. The axial component of this field, $\mathbf{B}(z,t)$, at the axis (i.e. $r = 0$) is obtained from the Biot–Savart law by~\cite{Jackson:1998nia}:
\begin{equation} \label{eq:mag-field-gen}
 \mathbf{B}(z,t)=\rm n_p \frac{\mu_0 I_p(t) R_p^2}{2\left(z^2+R_p^2\right)^{3 / 2}} \mathbf{e}_z,
\end{equation}
where $\rm z$ is the axial distance from the primary coil centre,  $\rm R_p$ the primary coil radius, $\rm n_p$ the number of turns, $\rm I_p(t)$ the excitation current, and $\mu_0$ the vacuum permeability. For a sinusoidal current, $\rm I_p(t) = I_0sin(\omega t)$ with amplitude $\rm I_0$ (typically on the order of 20~mA) and angular frequency $\omega$ ($\omega = 2\pi f$, with $f$ typically on the order of 10~kHz), the resulting magnetic flux $\Phi_{B}$ oscillates at the same frequency. The relation between this magnetic flux and magnetic field is~\cite{Jackson:1998nia}: 
\begin{equation}
\Phi_B(z, t)=\int_{\Sigma} \mathbf{B}(z, t) \cdot d \mathbf{S}
\end{equation}
with $\Sigma$ being the boundary of the outer coil and $d \mathbf{S}$ denoting the infinitesimal surface element perpendicular to the integration surface. According to Faraday's law of induction~\cite{Griffiths:1492149} this creates a voltage difference in both the outer coils:
\begin{equation}
\mathcal{E}=-\frac{d \Phi_B(z, t)}{d t}
\end{equation}
As the outer coils are counter-wound the difference between the two induced voltages, $V_{\rm diff} = V_{\rm out1} - V_{\rm out2}$, vanishes if the inductance on both coils is exactly the same. This happens when the source of the induction, i.e. the inner coil, is at the centre of the outer coils. Any small change in the position of the inner coil results in unequal induced voltages on the outer coils and a non-vanishing voltage difference $V_{\rm diff}$ is observed. This dependence of $V_{\rm diff}$ on primary coil position is very linear, which is one of the key features of an LVDT sensor. As the outer coils are connected in series $V_{\rm diff}$ is directly measured: its amplitude represents the absolute offset from the central position, while its phase determines the sign.

\begin{figure}[ht]
    \centering
    \begin{minipage}{0.49\linewidth}
        \includegraphics[width=\linewidth]{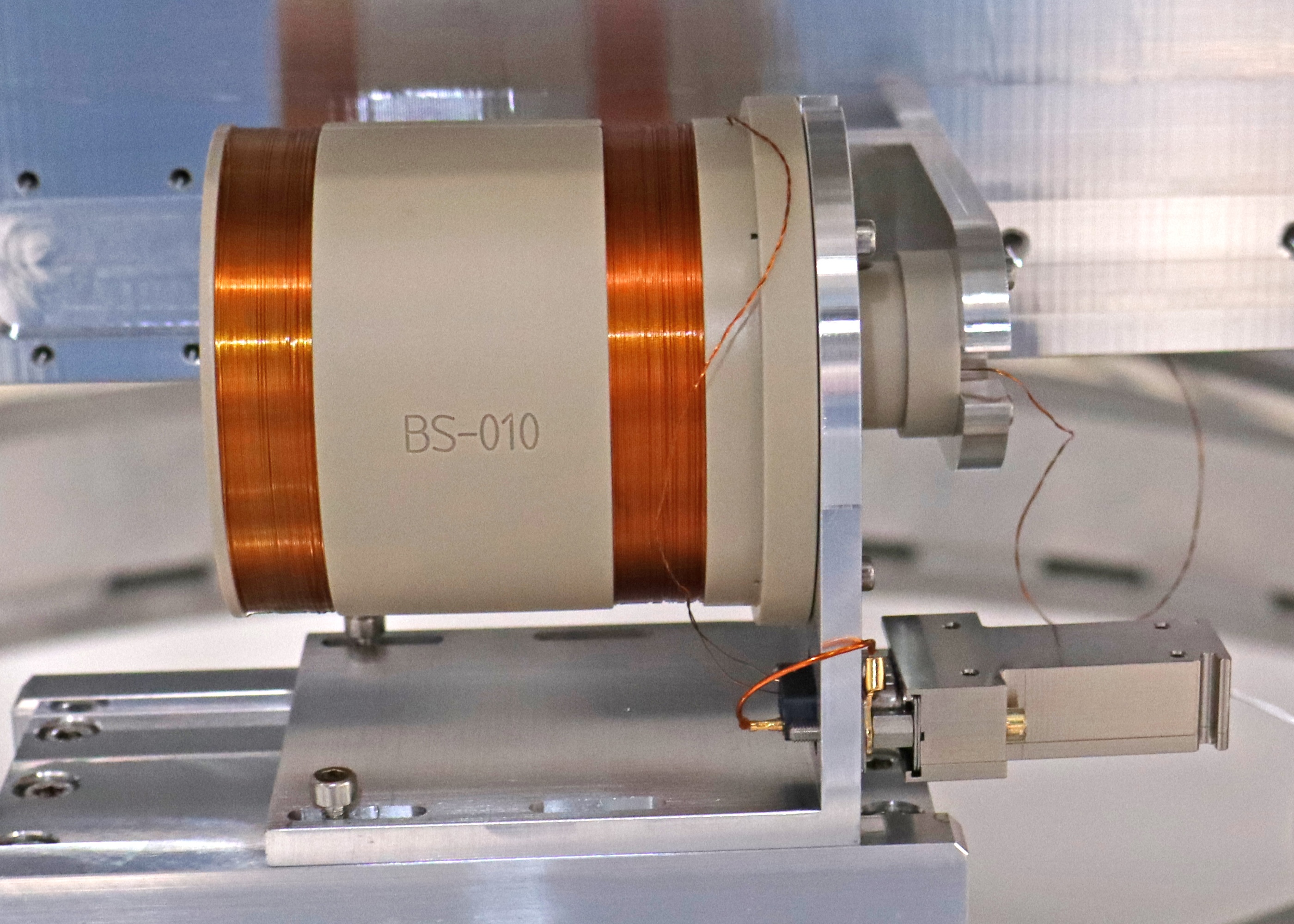}
    \end{minipage}
    \begin{minipage}{0.49\linewidth}
        \includegraphics[width=\linewidth]{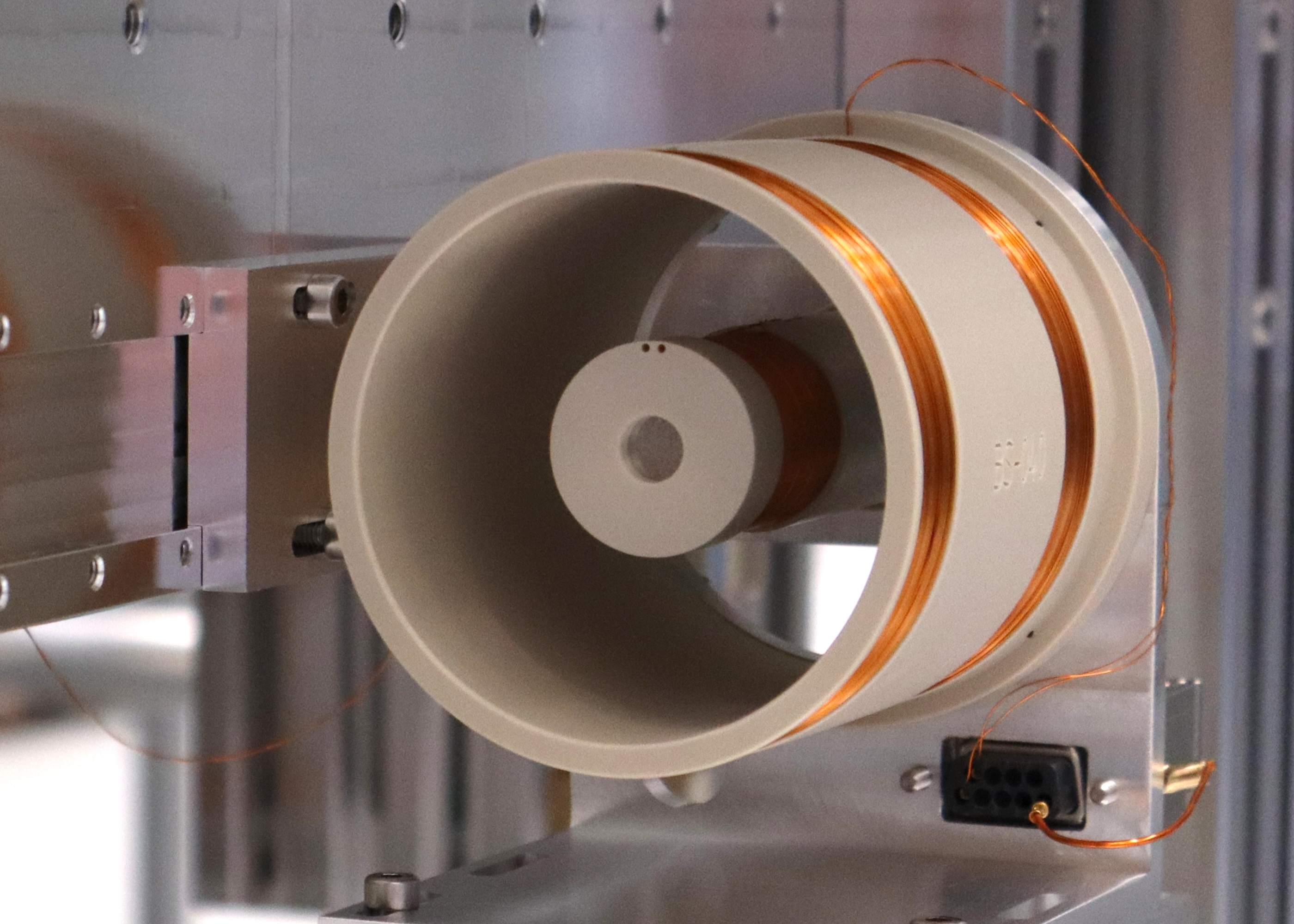}
    \end{minipage}
    \caption{Type-A LVDT+VC assembly installed in a seismic isolation stage of ETpathfinder~\cite{ETpathfinderTDR}. Left: clear view of the fixed, counter-wound secondary coils. Right: primary coil and permanent-magnet housing, illustrating the radial clearance between the primary and secondary coils that accommodates residual transverse suspension motion.}
    \label{fig:lvdt_vc}
\end{figure}

When a magnet is placed inside the primary coil, as shown in figure~\ref{fig:lvdt geometry}, and a DC current is applied to the secondary coils the system can generate a VC actuation force. This VC force is due to the Lorentz force exerted on a current carrying conductor, i.e secondary coil wire, in the magnetic field of the permanent magnet. The total force $\rm \mathbf{F}$ generated on the secondary coils can be expressed as~\cite{Jackson:1998nia}:
\begin{equation}
    \rm \mathbf{F} = n_s I_s\int_{L}(\mathbf{B} \times d\mathbf{l}),
\end{equation}
where $\rm I_s$ represents the current through the secondary coils, $\rm L$ is the length of the coil wire of one turn, $\rm d\mathbf{l}$ is the differential length vector along the coil, and $\rm n_s$ is the total number of turns. Due to the counter-wound two secondary coil geometry the maximum actuation force is obtained when the primary coil (or magnet) is at the central position, and decreases slightly when moving away. Thus, adding a permanent magnet to the LVDT geometry creates an integrated VC actuator so unwanted movements of the suspended mass can be corrected by providing the necessary force. 
Figure~\ref{fig:lvdt_vc} shows a real example of a combined LVDT+VC system, more specifically the Type-A configuration used in ETpathfinder that is characterised in this paper. It clearly shows the large difference between the radii of the primary and secondary coils. This design allows the suspended object to move freely in both the axial and transverse degree of freedom. All parameters and dimensions are summarised in table~\ref{tab:typeA_dimensions} and illustrated accordingly in figure~\ref{fig:lvdt geometry}. 

\begin{table}[ht]
    \centering
    \caption{Design parameters of an ETpathfinder Type-A LVDT+VC combination~\cite{ETpathfinderTDR}.}
    \label{tab:typeA_dimensions}
    \smallskip
    \begin{tabular}{|l|l|}
        \hline
        Parameter & Value\\
        \hline
        Secondary coil distance & 54.5 mm\\
        Secondary coil radius & 35.0 mm\\
        Secondary coil height & 13.5 mm\\
        Secondary coil layers & 7\\
        Primary coil radius & 11.0 mm\\
        Primary coil height & 24.0 mm\\
        Primary coil layers & 6\\
        Magnet radius & 5.0 mm\\
        Magnet height & 40.0 mm\\
        Magnet type & NdFeB N40\\
        Coil wire diameter & 32 AWG\\
        \hline
    \end{tabular}
\end{table}

To model this LVDT+VC system we use Finite Element Method Magnetics (FEMM) \cite{femm}, and its \texttt{pyFEMM} \cite{pyfemm} extension, to conduct two-dimensional axisymmetric simulations in the $r-z$ plane, where $r$ represents the radial direction and $z$ the vertical direction. To model the ambient environment, the simulation domain is divided into two radial regions: a dense mesh (element size 0.5) is used within 150~mm, while from 150~mm to 300~mm the mesh size is set automatically by FEMM, as beyond 150~mm the magnetic field has decayed sufficiently for a coarser discretisation to have negligible impact on the results. Dirichlet boundary conditions are set at the edges with a null vector potential. The LVDT coils and VC magnet are then positioned in this plane according to the ETpathfinder Type-A geometry, as summarised in table~\ref{tab:typeA_dimensions}. 
For the LVDT simulations, the primary coil is excited with a sinusoidal current of 20~mA at 10~kHz, reflecting the operating conditions of the electronics in the experimental setup. At this excitation level, the dissipated power ($\rm P$) is $\rm 4.9 \ mW$, obtained using the relation: $\rm P = I^2_{rms}R$, where $\rm I_{rms} (={20 \ mA} / \sqrt{2})$ is the root‑mean‑square excitation current and $\rm R (= 24.5 \ \Omega)$ is the coil resistance. For the VC simulations, the secondary coils are driven with a 1~A DC current. In both cases, the axial position of the primary coil is varied relative to the secondary coils, and FEMM solutions are computed at discrete displacement steps. For the LVDT analysis, complex voltages are extracted from the FEMM secondary-coil circuit properties and used to calculate the absolute differential output voltage. For the VC analysis, the actuation force is determined from the axial component of the force (calculated using the steady-state weighted Maxwell stress tensor~\cite{pyfemm}) on the secondary coils. This procedure results in a set of discrete data points representing the sensor output (differential voltage) or actuator response (axial force) as a function of relative coil position. From these data, an LVDT response profile is constructed, with the displacement of the primary coil relative to the secondary coils on the x-axis and the differential voltage on the y-axis. Similarly, a VC force profile is constructed, where the axial force is plotted as a function of displacement. These profiles provide a quantitative characterisation of the system, and are used to extract three key metrics to evaluate the ETpathfinder Type-A assembly: the LVDT response, LVDT linearity, and normalised VC force, each described below.

The LVDT response $(\rm V_{out}/mmA_{in})$ is defined as the slope of a first-order polynomial fit to the differential voltage of the secondary coils ($\rm V_{out}$) as a function of the primary coil position. This slope is normalised to the input excitation current ($\rm A_{in}$) or to the excitation voltage ($\rm V_{in}$) to compare with measurements. The fit range is smaller than the full simulated or measured position range, and is restricted to a central interval in order to avoid bias from nonlinearities in the LVDT output at larger displacements. This effect is quantified by the LVDT linearity (in \%): the degree to which the differential output voltage follows a linear dependence on the primary coil position. A high linearity is essential for precision displacement sensing, as deviations from an ideal linear behaviour introduce systematic errors in the inferred position. The linearity is calculated as a relative fit residual:
\begin{equation}
L({\rm z}) = \rm 100\times \frac{\lvert V_{fit}(z) - V_{data}(z)\rvert}{\lvert V_{data}(z) \rvert},
\end{equation}
where $\rm V_{data}$ denotes the simulated or measured differential voltage, $\rm V_{fit}$ is the value obtained from the linear fit, and z represents the corresponding primary coil position. Finally, the normalised VC force (N/A) is defined as the axial actuation force, $\rm F(z)$, generated by the voice coil as a function of primary coil position z, normalised to the applied current in the secondary coils.
\section{Experimental setup} 
\label{sec:exp-setup}

\begin{figure}[ht]
    \centering
    \includegraphics[width=.555\linewidth]{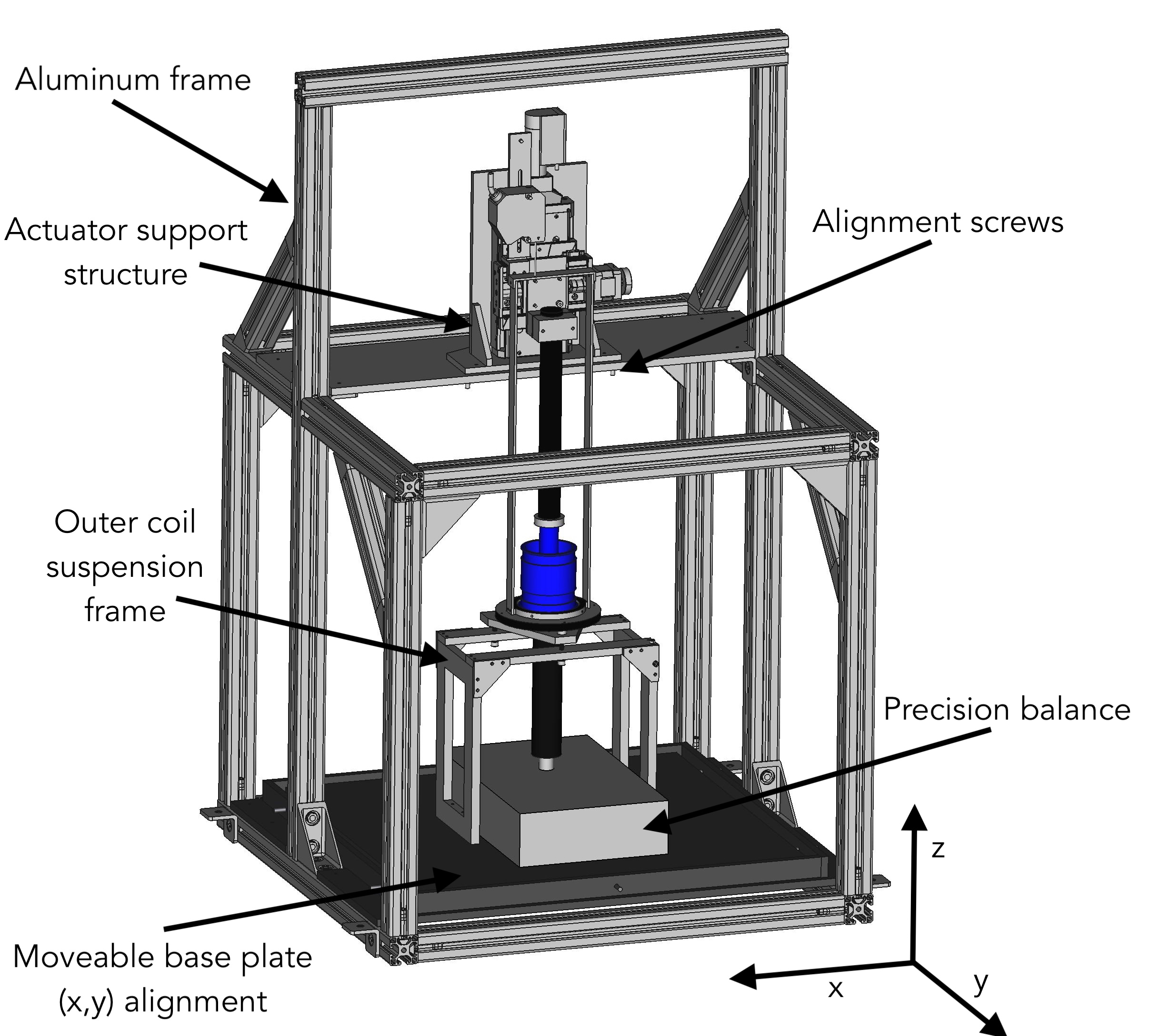}
    \includegraphics[width=.435\linewidth]{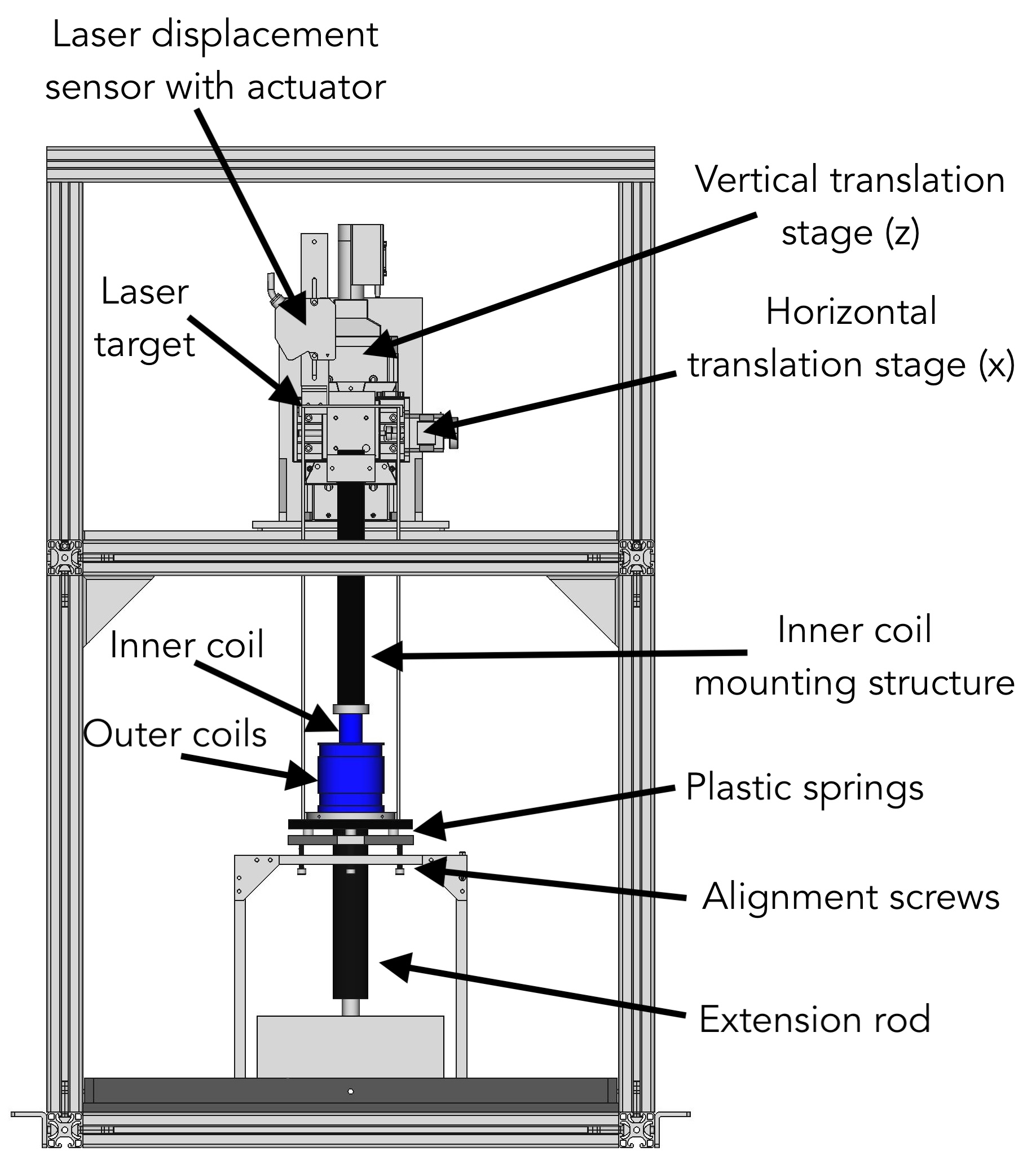}
    \caption{Drawing of the measurement setup at the University of Antwerp, capable of characterising both LVDT sensors and VC actuators. A 3D view (left) and front view (right) provide an overview with all major components indicated. }
    \label{fig:exp-setup}
\end{figure}

To study and develop novel LVDT sensors and VC actuators for GW detectors a dedicated measurement setup was constructed at the University of Antwerp to precisely characterise the performance of assembled prototypes. It is used to experimentally measure the LVDT response, LVDT linearity, VC actuation force, and VC force stability of the ETpathfinder Type-A configuration, and compare the data with obtained simulation results. The complete setup is presented in figure~\ref{fig:exp-setup} and described below.

The entire system is housed within an aluminium frame to ensure stability, and is placed on a passive suspension table that suspends the setup from high-frequency vibrations. At the bottom, a movable base plate is present to align the outer coil with respect to the inner coil in the $(x,y)$ plane. Two linear translation stages are mounted on the actuator support structure and control the motion of the installed primary (inner) coil of the LVDT or VC. Along the $x$-axis a VT-80-62309220 stage from Physik Instrumente (PI) SE \& Co. KG~\cite{physikinstrumente_website} is used to measure at different transverse inner coil positions. It contains a 2-phase stepper motor, has a resolution of $5~\mu$m, a minimal step size of $0.2~\mu$m, unidirectional repeatability of $0.4~\mu$m, and a travel range of 50~mm. Along the $z$-axis a PI M-406.2DG stage is used to precisely set the axial positions of the inner coil. It has a DC servo motor, provides a resolution of $8.5$~nm, minimal step size of $0.1~\mu$m, unidirectional repeatability of $0.2~\mu$m, and a travel range of 50~mm. The settings and motion of the stages is remotely controlled using \texttt{PIPython}~\cite{pipython_2.11.0.6}. The inner coil is attached to the stages via a 285~mm long plastic extension bar, which allows for adjustable mounting positions to accommodate various designs. 

In addition, a laser reference measurement system is present, and consists of a laser displacement sensor and actuator. It measures the true distance between the inner and outer coils along the $z$-axis through an extension structure (with laser target) from the outer coil holder with high precision, serving as a reference to confirm the motion generated by the translation stages. The laser displacement sensor is a Keyence LK-H052~\cite{keyence_lk-g5000}. It has a reference distance of 50 mm, a measurement range of $\pm10$~mm, and a resolution of 25~nm. The settings of the laser sensor, as well as data acquisition, are controlled via the RS-232C protocol. To extend its measurement range it is mounted in a sliding slot and attached to a L12-I micro linear actuator~\cite{actuonix_L12-50-50-12-I}, which has a full stroke of 50~mm and a resolution of 0.3~mm.

To measure the force generated by a VC actuator a spring-balance system is mounted at the bottom of the setup. It consists of a plastic outer coil suspension frame and a precision balance (OHAUS PX224~\cite{ohaus_px224_balance}, with RS-232 readout), which has a capacity of 220~g and a resolution of 0.1 mg. The outer coil is mounted on a support platform (black in figure~\ref{fig:exp-setup}) that is suspended by three plastic springs. These are supported by a structure that can be adjusted with alignment screws. Through a 190~mm extension rod the weight of this system is transferred to the balance, and can be tuned with the springs and alignment screws to ensure it is below 220~g regardless of the weight of the installed VC prototype. As the coils are mounted vertically, the balance measures the force generated by the VC as a change in weight ($\Delta m$), which is then converted into force via $\mathbf{F} = k\Delta m\mathbf{g}$, with $\rm \mathbf{g} = 9.81\ m/s^2$ the gravitational acceleration, and $k$ a correction factor to take the mechanical coupling of the springs into account. The plastic springs act as a damping system, which allows the VC actuator to produce low-frequency oscillations that are used to measure the force. 

It is important to note that no ferromagnetic or conductive metallic materials are used in the construction of mounting or support structures within a radius of 300~mm around the LVDT and VC coils. This design choice minimises interference from eddy currents and stray magnetic fields in the vicinity of the coils.

To fully control all components of the setup and automate measurements, we developed a custom \texttt{python} software package based on \texttt{PIPython} and \texttt{pyserial}~\cite{pyserial}. This package allows us to manage the motion of both translation stages, acquire real-time data from the laser sensor and precision balance, and communicate with the data acquisition (DAQ) system for the digital signal processing stage. The DAQ system used with this experimental setup is developed by the Laboratoire d’Annecy de Physique des Particules (LAPP)~\cite{virgo_daq_lapp}, and is exactly the same as the DAQ of the Advanced Virgo GW detector and ETpathfinder. This facilitates studies of the LVDT or VC signal readout, and the development of novel amplification electronics. It recreates the GW detector signal processing and allows us to apply our measurement results directly to ETpathfinder. The signal is digitised via an ADC module (18-bit, 1~MHz sampling rate, differential voltage span $\pm 10$~V) with a noise level of 90~nV/$\rm \sqrt{Hz}$ above 100~Hz, and can then be digitally demodulated or stored for further analysis. To generate LVDT excitation or VC actuation signals a DAC module is available (24-bit, 400~kHz sampling rate, dynamic range $\pm 10$~V with maximum current $\pm 20$~mA) with a noise level of 100~nV/$\rm \sqrt{Hz}$ above 30~Hz.

However, before signals are sent to, or received from, the DAQ system they pass through a dedicated analog electronics circuit, developed for the Type-A LVDT+VC combination for Advanced Virgo and ETpathfinder. This contains an amplification section that increases the LVDT output to obtain the required SNR along with common-mode noise reduction, a circuit to provide the primary coil excitation voltage, and a VC driver section to generate low-frequency current signals from DAC voltage outputs. Through a series of instrumentation and operational amplifiers the circuit is optimised to amplify differential LVDT signals of 10~kHz, and a total gain factor of 65, 260, or 850 can be configured. The VC driver circuit is designed to reduce the control noise of the DAQ, and has a low-pass filter with 20~dB attenuation at 2.8~Hz. A high output-impedance current source converts the voltage input and creates a stable low-frequency current signal. Lastly, to enable a systematic study of the individual signal-processing stages and to compare the raw LVDT response with the amplified and digitised response, a Tektronix DPO4054 oscilloscope~\cite{man-tektronix_dpo4054} with a maximum sampling rate of 250~MHz is used to directly measure the differential output of the secondary coils.
\section{LVDT measurements} 
\label{sec:LVDT}

To measure the LVDT response the primary and secondary coils are mounted on their respective support structures, and carefully aligned to minimise any transverse offsets and ensure that all coils are perfectly levelled. The primary coil is excited with a sinusoidal signal ($f = 10~{\rm kHz}$) using a DAC module of the DAQ system, and two independent measurements were performed to record the differential voltage output of the secondary coils: one with the oscilloscope connected (pre-amplification) and another with DAQ system connected (post-amplification). For the former the amplitude of the excitation signal is set to $V_{\rm in} = 9$~V, while for the latter it is set to $V_{\rm in} = 0.1$~V\footnote{These voltages are set based on the available signal amplification and dynamic range of the ADC of the data acquisition system}. The first step in the measurement procedure is to locate the central position of the primary coil by scanning a large range in $z$ with the vertical translation stage and find the coordinate $z_0$ where the LVDT output becomes minimal. This is determined with a precision of $10~\mu$m and provides a reference for the actual measurement range. The primary coil is then displaced within $\pm 5$~mm from its central position with discrete steps of 0.5~mm, and at each position the secondary coils output is recorded. The time-domain data are fitted with a sine wave function, $V(t) = V_{\rm out}\sin(\omega t+\phi) + c$, to extract the amplitude $V_{\rm out}$ at each position. The phase ($\phi$) and bias ($c$) parameters are free, while the frequency parameter is fixed to 10~kHz. The normalised LVDT output is then obtained by dividing all $V_{\rm out}(z)$ by $V_{\rm in}$ and plotting these results as a function of axial displacement relative to the central primary coil position. With a linear polynomial fit, $V_{\rm out}(z)/V_{\rm in} = s(z-z_0)+b$, the slope parameter $s$ can be determined to obtain the LVDT response~([V/mmV]), and the LVDT linearity can be calculated as defined in section~\ref{sec:working_principle}.  

\begin{figure}[ht]
    \centering
    \begin{minipage}{0.49\linewidth}
        \includegraphics[width=\linewidth]{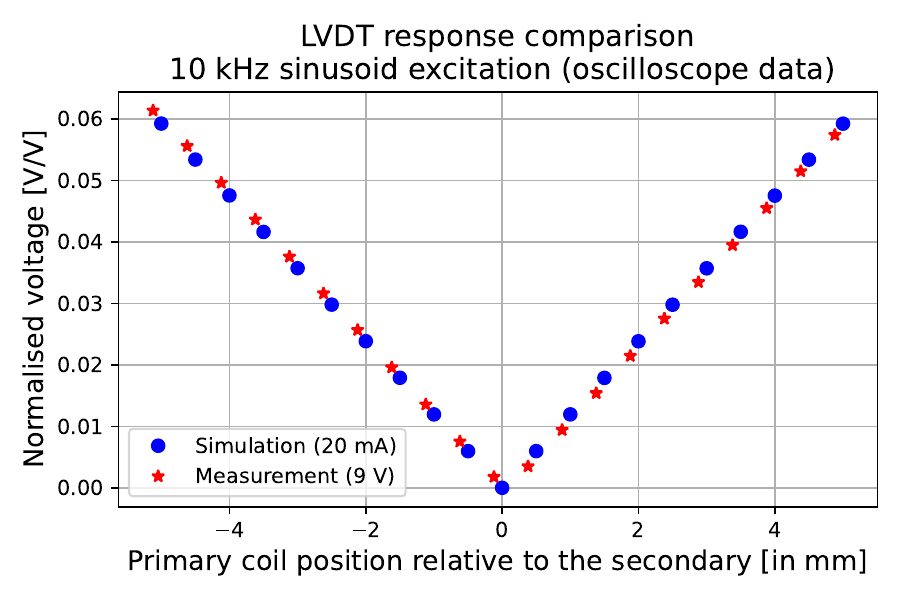}
    \end{minipage}
    \begin{minipage}{0.49\linewidth}
        \includegraphics[width=\linewidth]{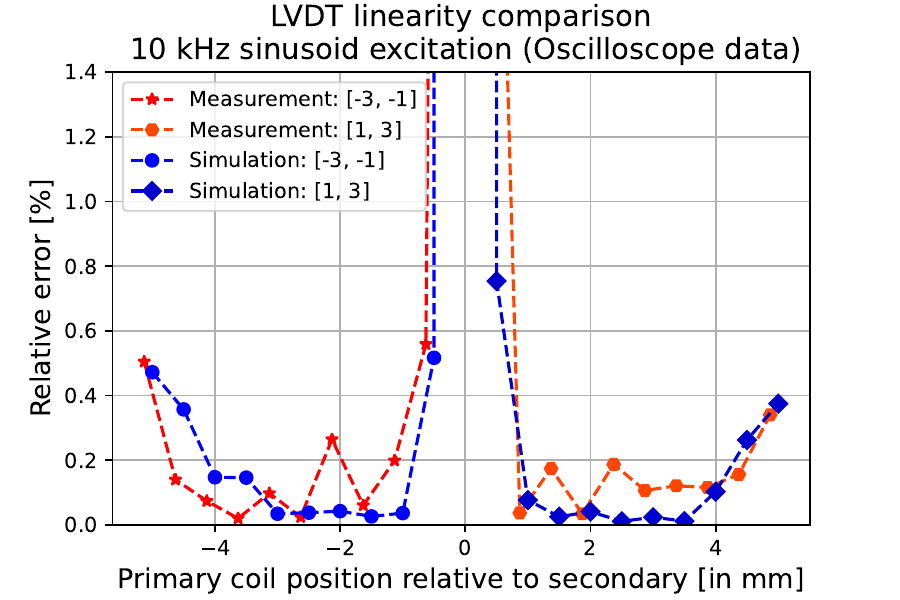}
    \end{minipage}
    \caption{Measurement of the normalised LVDT output (left) and linearity (right) in a $\pm 5$~mm range (step size 0.5~mm) with the oscilloscope (pre-amplification), compared to FEMM simulations. The primary coil is excited with 9~V (data) and 20~mA (simulation) 10~kHz sine wave. The data points are fitted with a linear polynomial in $[-3,-1]$~mm and $[1,3]$~mm to extract the LVDT response and plot the linearity.}
    \label{fig:scope response}
\end{figure}

Figure~\ref{fig:scope response} shows the normalised LVDT output (left) and the obtained linearity (right) when measured with the oscilloscope. The FEMM simulation results (blue) show a good agreement with the data points\footnote{These are slightly shifted in relative position due to $z_0$ not exactly coinciding with a 0.5~mm step.} (red) in both observables. However, due to the low pre-amplification output signal and the presence of electronic noise the linear behaviour of the LVDT is lost within a region of $\pm0.5$~mm around the centre $z_0$. To avoid any bias of the non-linear regions both data and simulation are therefore independently fitted over the negative and positive displacement domains $[-3,-1]$~mm and $[1,3]$~mm respectively. The final LVDT response is then the mean value of both fit results. Nevertheless, even though the linearity decreases when going past $\pm 4$~mm the relative error of the fit remains well below 1\%, which confirms that the measured Type-A LVDT+VC combination can be used as a precise position sensor with a dynamic range of $\pm 5$~mm.

\begin{figure}[ht]
    \centering
    \begin{minipage}{0.49\linewidth}
        \includegraphics[width=\linewidth]{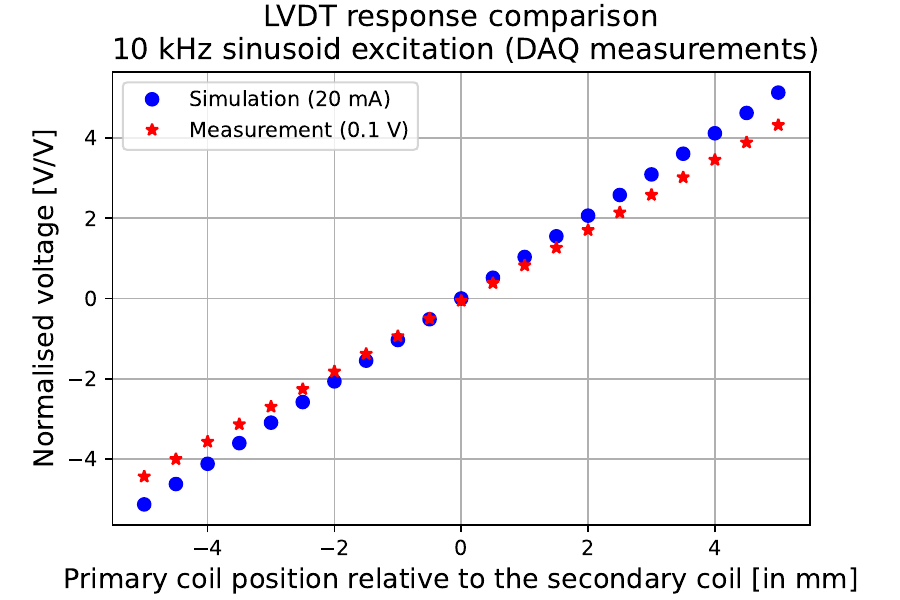}
    \end{minipage}
    \begin{minipage}{0.49\linewidth}
        \includegraphics[width=\linewidth]{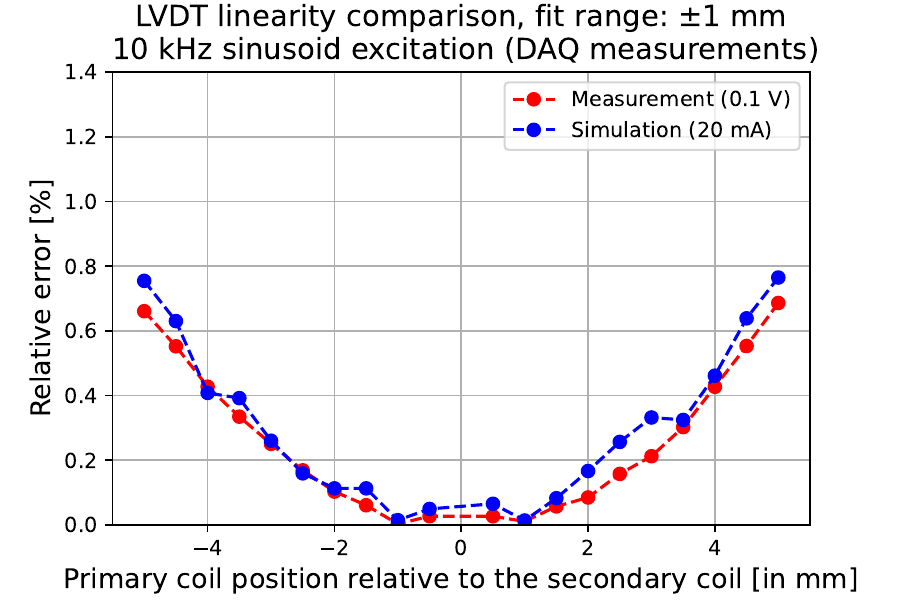}
    \end{minipage}
    \caption{Measurement of the normalised LVDT output (left) and linearity (right) in a $\pm 5$~mm range (step size 0.5~mm) with the DAQ system (post-amplification), compared to FEMM simulations scaled with the electronics gain factor. The primary coil is excited with a 0.1~V (data) and 20~mA (simulation) 10~kHz sine wave. The data points are fitted with a linear polynomial within $[-1,1]$~mm to extract the LVDT response and plot the linearity.}
    \label{fig:lvdt response}
\end{figure}

The post-amplification measurements with the DAQ system are shown in figure~\ref{fig:lvdt response}. The normalised LVDT output (left) and linearity (right) are now compared to FEMM simulations that are scaled with the total gain factor of the electronics circuit. Due to the high output signal, electronic filtering, and common-mode noise rejection, the linearity is preserved around $z_0$, and it is sufficient to fit the data points with one linear function within $[-1,1]$~mm to extract the LVDT response and minimise any bias from non-linear behaviour at larger position offsets. 
The measured LVDT linearity agrees well with the simulation and shows a smooth increase of the relative fit error. At $\pm 5$~mm this remains below 0.8\%, which again confirms that the considered LVDT can be used as a highly linear position sensor within this dynamic range. However, a significant discrepancy is found between the simulated and measured LVDT output. This is attributed primarily to the electronic components responsible for amplifying the LVDT signal\footnote{The measured gain of the instrumentation amplifiers was found to be lower than the theoretical gain.}. To resolve this deviation, a data-driven correction factor will be determined and applied to simulations, incorporating all relevant systematic uncertainties as explained in the following section.

\section{Uncertainty analysis} 
\label{sec:systematics}

To quantify the reliability of the measurements and to compare them meaningfully with simulations, both statistical and systematic uncertainties are considered for all results in this paper. The statistical uncertainties ($\sigma_{\rm stat}$), only relevant for experimental data, are calculated from repeated measurements under identical conditions, and reflect random noise and other non-systematic variations. For both LVDT and VC measurements these are found to be small compared to the systematic contributions. The systematic uncertainties ($\sigma_{\rm sys})$ arise from controlled variations of key experimental and simulation parameters. In addition, for each systematic source upper and lower deviations ($\sigma^+$, $\sigma^-$) are evaluated. To study the contribution of each systematic effect separately its uncertainty including the statistical contribution is computed in quadrature:
\begin{equation}
    \sigma^{\pm}_{\rm sys, tot} = \sqrt{\sigma^{2}_{\rm stat} + \left(\sigma^{\pm}_{\rm sys}\right)^{2}}.
\end{equation}
The overall uncertainty is then obtained by combining the statistical term and all systematic sources:
\begin{equation}
    \sigma^{\pm}_{\rm tot} = \sqrt{\sigma^{2}_{\rm stat} + \sum (\sigma^{\pm}_{\rm sys})^{2}}.
\end{equation}
The considered systematic uncertainties on the LVDT response arise from the following four sources: 
\begin{itemize}
    \item \textbf{Excitation voltage ($V_{\rm in}$)}: directly scales the sensor response and may lead to residual effects for low values of $V_{\rm in}$. In addition, the choice of the actual value depends on the used DAQ system and sensor application. Therefore we modified the default excitation value of the oscilloscope measurements (9~V) with $\pm 0.5$~V to study the stability of the response after voltage normalisation. For the DAQ measurements the default value (0.1~V) was modified to 0.05~V and 0.2~V.
    \item \textbf{Transverse offset}: reflects potential transverse coil misalignments in the experimental measurement setup. This affects the symmetry of the magnetic field and hence the coupling. Therefore the primary coil transverse position was modified with $\pm 1$~mm.  
    \item \textbf{Step size}: refers to the measured or simulated positions of primary coil displacement, and influences the linear fit results. To study how the sensor response is affected by this choice we modified the default step size (0.5~mm) to 0.25~mm and 1.0~mm.
    \item \textbf{Mesh resolution}: determines the accuracy of the field gradients in the simulation model. Coarse mesh sizes can lead to inaccurate predictions, while very fine meshes increase computational cost without necessarily improving accuracy. The default setting (0.5) of Airspace-2 was adjusted to 0.1 (very fine grid) and 1.0 (coarse grid) to study the influence on LVDT response.
\end{itemize}
The excitation voltage and transverse offset systematics are only applied to data, while the mesh resolution systematic applies only to simulations. A more detailed study of these systematics is provided in \cite{kumar-thesis}. In the following, these systematics are combined into a quantitative uncertainty budget for both oscilloscope and DAQ-based measurements. In all plots, the error bars represent the maximum positive and negative deviations associated with each systematic parameter.

\subsection{Systematic uncertainties on oscilloscope measurements} 
\label{subsec:osc_uncertainties}

\begin{figure}[ht]
    \centering
    \begin{minipage}{0.49\linewidth}
        \includegraphics[width=\linewidth]{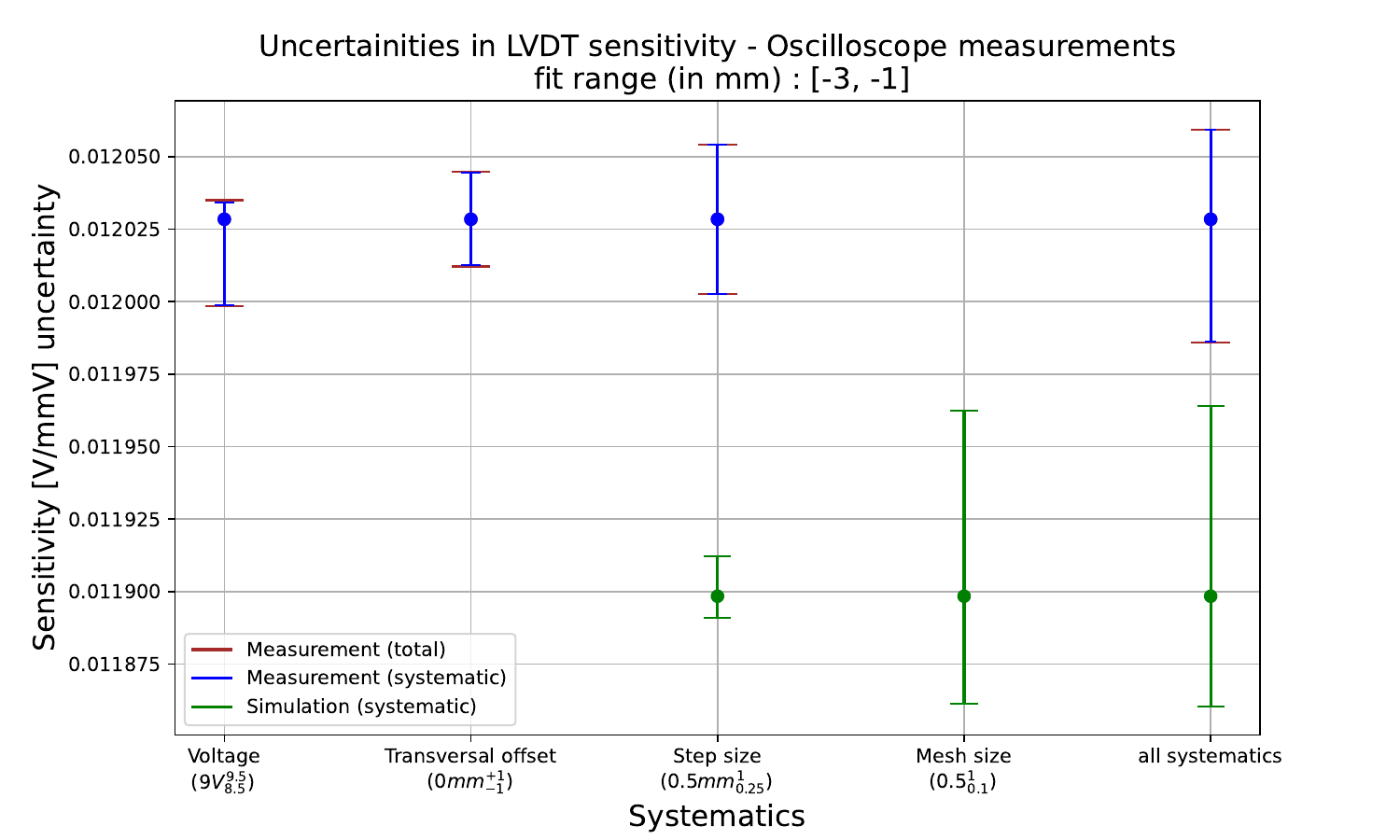}
    \end{minipage}
    \begin{minipage}{0.49\linewidth}
        \includegraphics[width=\linewidth]{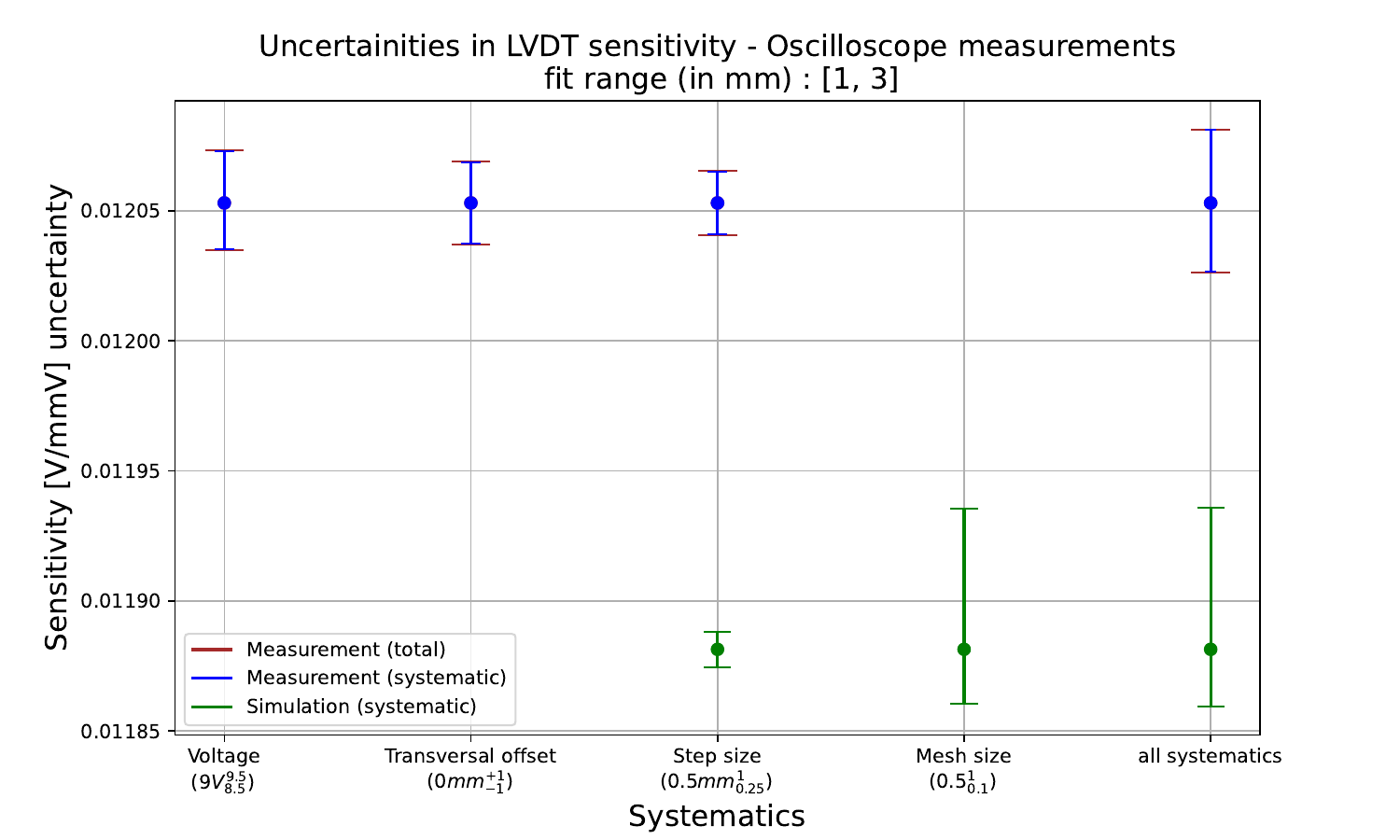}
    \end{minipage}
    \caption{Uncertainty analysis for oscilloscope-based measurements and FEMM simulations. The normalised LVDT responses [V/mmV] are obtained from independent fits to the negative $[-3,-1]$~mm (left) and positive $[1,3]$~mm (right) displacement ranges. The blue (green) error bars represent the individual systematic contributions in the measurement (simulation), the red error bars also include the statistical uncertainty of the measurement. }
    \label{fig:lvdt scope uncertainties}
\end{figure}

For oscilloscope-based measurements, the response is fitted separately on the negative $[-3,-1]$~mm and positive $[1,3]$~mm sides of the displacement range; consequently, the systematic uncertainties are analysed independently before being combined into a single mean LVDT response. As shown in figure~\ref{fig:lvdt scope uncertainties}, systematic uncertainties dominate over the statistical error. On the negative side (left), the excitation voltage variation contributes between $-0.25\%$ and $+0.05\%$, the transverse offset contributes $\pm0.14\%$, and the step-size uncertainty contributes $\pm0.22\%$. On the positive side (right), the corresponding contributions are $-0.15\%$ to $+0.17\%$ from the excitation voltage, $\pm0.13\%$ from the transverse offset, and $\pm0.10\%$ from the step size. The excitation voltage thus shows the largest measurement uncertainty, while the contribution from transverse offset is consistently smaller. For the simulations the mesh resolution dominates, in the negative range its systematic uncertainty contributes $-0.31\%$ to $+0.54\%$, while the step-size contributes between $-0.06\%$ to $+0.12\%$. In the positive range, the mesh-size contributions are $-0.18\%$ to $+0.46\%$, and the step-size contributes $\pm0.06\%$.

\begin{figure}[ht]
    \centering
    \includegraphics[width=0.8\linewidth]{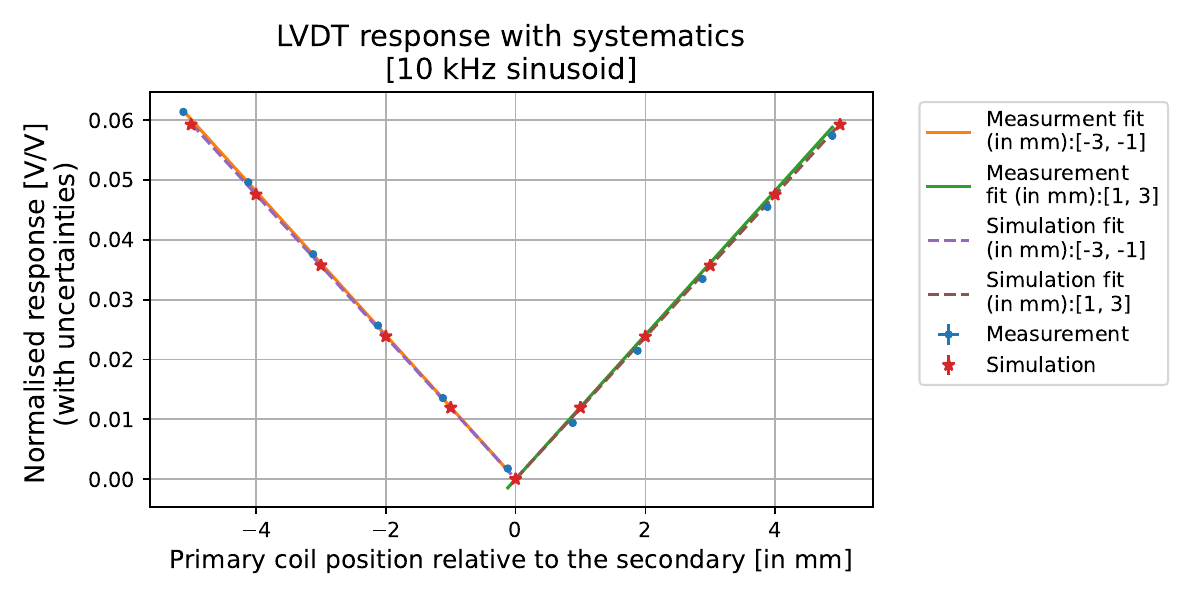}
    \caption{Oscilloscope and simulation data, together with independent linear fits on the negative $[-3,-1]$~mm and positive $[1,3]$~mm displacement ranges. The data points are plotted with a 1~mm step size, as only those can be evaluated with all systematic contributions. The error bars represent the total uncertainties but are too small to be visible.}
    \label{fig:lvdt scope fits bothsides}
\end{figure}

\begin{figure}[ht]
    \centering
    \includegraphics[width=0.6\linewidth]{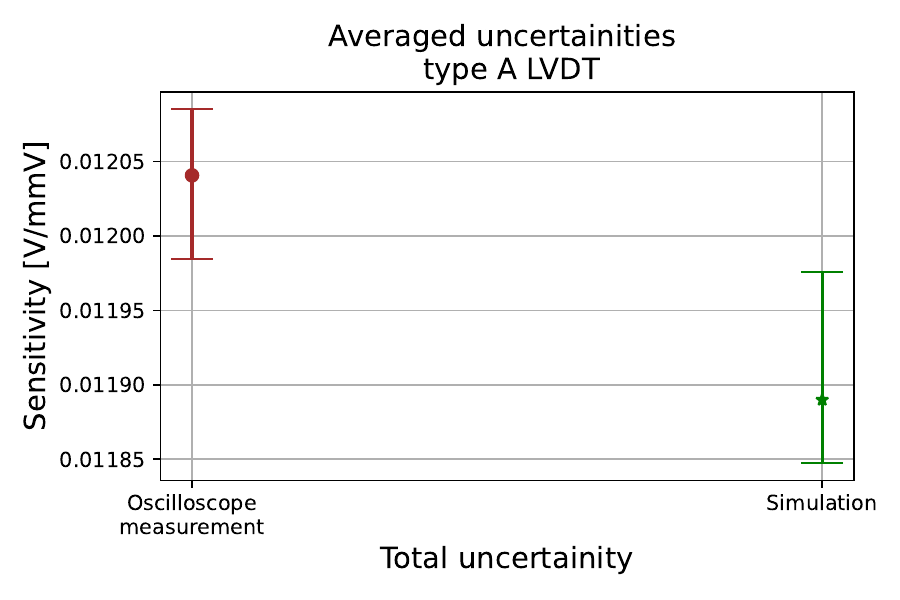}
    \caption{Comparison of the average LVDT response obtained from oscilloscope-based measurements and FEMM simulations. The error bars denote the combined statistical and systematic uncertainties.}
    \label{fig:average errorbar}
\end{figure}

Figure~\ref{fig:lvdt scope fits bothsides} shows the measurement and simulation data, together with their independent linear fits on the negative and positive displacement ranges. The slope parameters obtained from these two fits (left and right) are then averaged to provide a single normalised LVDT response value for the oscilloscope measurement and FEMM simulation. These values, together with their combined uncertainties, are shown in figure~\ref{fig:average errorbar}.

The average measured LVDT response is $s_{\rm osc} = 0.01204^{+0.00004\ (+0.37\%)}_{-0.00006\ (-0.46\%)}$~V/mmV, while the FEMM simulation predicts a response of $s_{\rm sim} = 0.01189^{+0.00009\ (+0.73\%)}_{-0.00004\ (-0.36\%)}$~V/mmV. The experimentally determined value exceeds the simulated prediction by only $1.3\%$, demonstrating that the FEMM model correctly calculates the LVDT response. The close agreement between measurement and simulation, together with the high precision and linearity of the measured response, confirms the stability and accuracy of the experimental setup. These results validate the ETpathfinder Type-A LVDT+VC combination as a highly linear and reliable position sensor over a displacement range of $\pm 5$~mm.

\subsection{Systematic uncertainties on DAQ measurements} 
\label{subsec:daq_uncertainities}

\begin{figure}[ht]
    \centering
    \includegraphics[width=0.65\linewidth]{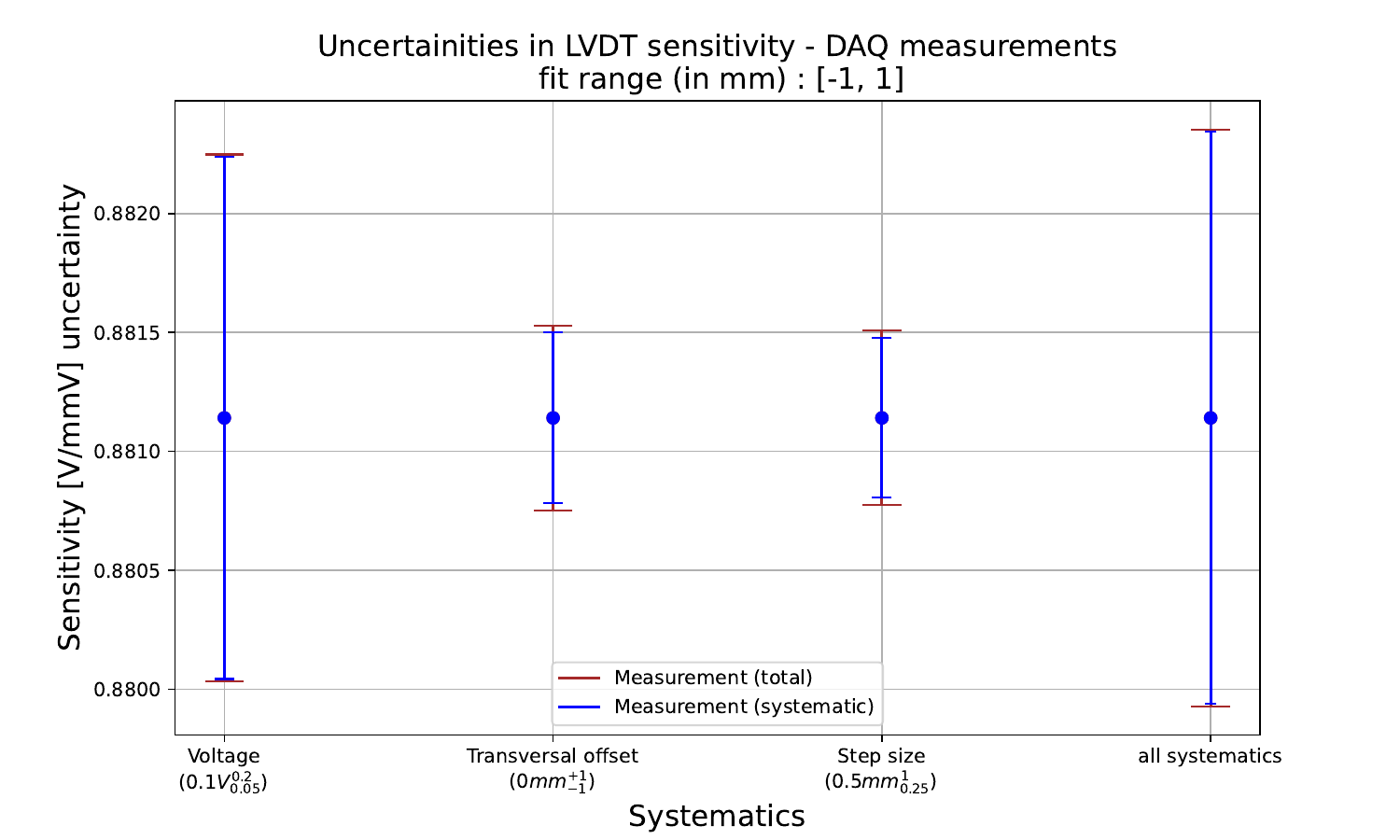}
    \caption{Uncertainty analysis for DAQ-based measurements. The normalised LVDT response [V/mmV] is obtained from a single fit in the $[-1,1]$~mm displacement range. The blue error bars represent the individual systematic contributions in the measurement, the red error bars also include the statistical uncertainty of the measurement.}
    \label{fig:daq uncertainities daq box}
\end{figure}

For DAQ-based measurements performed after signal amplification, a single linear fit within $[-1,1]$~mm is sufficient to determine the LVDT response. The corresponding uncertainty analysis, summarised in figure~\ref{fig:daq uncertainities daq box}, indicates that statistical uncertainties are again negligible, with the dominant systematic contribution arising from variations in the excitation voltage. The relative uncertainties associated with the excitation voltage, transverse offset, and step size are $\pm0.13\%$, $\pm0.04\%$, and $\pm0.04\%$, respectively. All systematic sources thus introduce symmetric errors. The measured LVDT response with total uncertainty is then found to be $s_{\rm daq} = 0.88114 \pm 0.00121\ (\pm 0.14\%)$~V/mmV.

\subsection{Data-driven correction factor}
\label{subsec:correction_factor}

As previously discussed, a significant discrepancy was observed between the DAQ measurements and the FEMM simulations in figure~\ref{fig:lvdt response}. This is mainly due to unmodelled effects in the electronic signal chain, which results in a net total gain that deviates from the expected value solely based on the amplifiers, filters, and circuit impedance. To account for these effects, a data-driven correction factor is defined as:
\begin{equation}
    C_{\rm ds} = \frac{s_{\rm daq}}{s_{\rm osc}} 
           = \frac{0.88114_{-0.00121}^{+0.00121}}{0.01204_{-0.00006}^{+0.00004}} 
           = 73.18_{-0.38}^{+0.26},
\end{equation}
with the uncertainties on the DAQ- and oscilloscope-derived LVDT responses considered to be uncorrelated. The correction factor effectively encapsulates all contributions of the real electronics chain into a single data-driven scaling parameter. Using this correction factor, the simulated LVDT response can be rescaled according to:
\begin{equation}
    s_{\rm corr} 
    = C_{\rm ds} \times s_{\rm sim} 
    = 73.18^{+0.26}_{-0.38} \times 0.01189^{+0.00009}_{-0.00004} 
    = 0.87011^{+0.00728}_{-0.00538}\,.
\end{equation}

\begin{figure}[ht]
    \centering
    \includegraphics[width=0.6\linewidth]{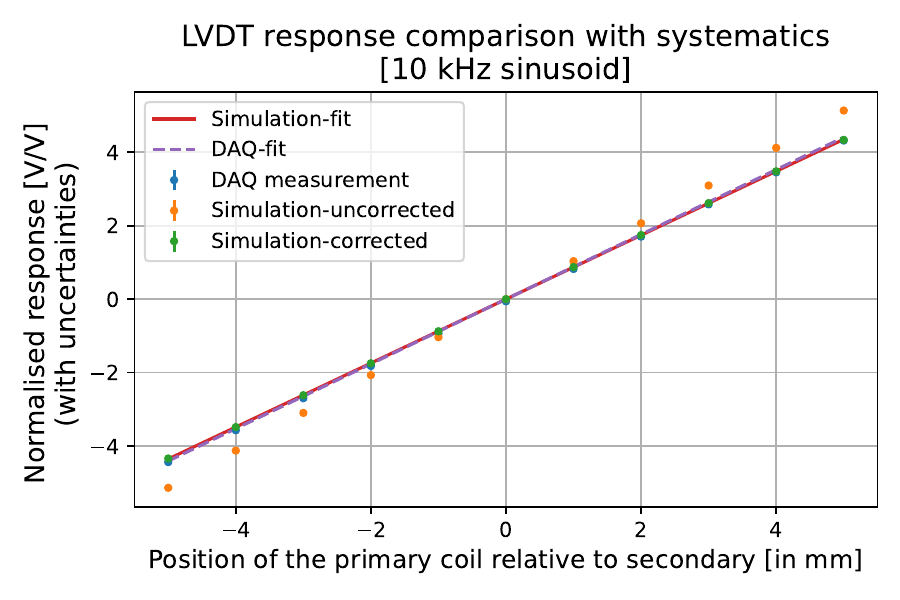}
    \caption{Comparison of the DAQ-based LVDT response measurement (blue) with the uncorrected simulation (orange) and $C_{\rm ds}$-corrected simulation (green) values. The results of the linear fits within $[-1,1]$~mm on data (blue dashed) and the corrected simulation (red) are shown over the full dynamic range. The LVDT response from the corrected simulation agrees with the DAQ measurement within 1.3\%. The error bars represent the total uncertainties, but are too small be to visible.}
    \label{fig:lvdt-corrected}
\end{figure}

This allows us to consistently compare all results of the Type-A assembly at the level of the DAQ system, and makes it straightforward to include future simulation models. In addition, the measured values can be directly applied in ETpathfinder or any other GW detector employing the same DAQ infrastructure. Figure~\ref{fig:lvdt-corrected} shows the $C_{\rm ds}$-corrected simulated LVDT response, including its associated uncertainties (green), compared with the DAQ measurements and their uncertainties (blue). The two responses agree very well, within 1.3\%, as was observed with the oscilloscope measurement. The fitted curves (red and dashed blue), obtained from the corrected simulation and the DAQ measurements, respectively, overlap almost entirely. This agreement demonstrates that the corrected FEMM model provides an accurate estimate of the post-amplification LVDT response, effectively eliminating the discrepancy between the expected and measured responses previously observed in figure~\ref{fig:lvdt response}. The complete uncertainty budget, and an overview of LVDT response values for all measurements and FEMM simulations are summarised in table~\ref{tab:lvdt_systematics_summary}.

\begin{table}[ht]
    \centering
    \caption{Complete uncertainty budget and overview of LVDT response values for all measurements and FEMM simulations. For each individual measurement type and systematic source the relative uncertainty variation is shown. The experimental measurements are most affected by excitation voltage, and the FEMM simulations by mesh size. The measured and simulated LVDT responses agree within 1.3\%.}
    \label{tab:lvdt_systematics_summary}
    \smallskip
    \renewcommand{\arraystretch}{1.25}
    \setlength{\tabcolsep}{7pt}
    \begin{tabular}{|c|c|c|c|c|c|}
        \hline
        \multirow{2}{*}{\textbf{Measurement}} & 
        \multirow{2}{*}{\textbf{Range (mm)}} & 
        \multicolumn{4}{c|}{\textbf{Systematic variations (\%)}} \\ 
        \cline{3-6}
         & & Voltage & Offset & Step size & Mesh \\
        \hline
        \multirow{2}{*}{Oscilloscope} 
            & [$-3$, $-1$] & $-0.25$ to $+0.05$ & $\pm0.14$ & $\pm0.22$ & -- \\ 
            & [1, 3] & $-0.15$ to $+0.17$ & $\pm0.13$ & $\pm0.10$ & -- \\
        \hline
        \multirow{2}{*}{Simulation} 
            & [$-3$, $-1$] & -- & -- & $-0.06$ to $+0.12$ & $-0.31$ to $+0.54$ \\
            & [1, 3] & -- & -- & $\pm0.06$ & $-0.18$ to $+0.46$ \\
        \hline
        \multirow{1}{*}{DAQ} 
            & [$-1$, 1] & $\pm0.13$ & $\pm0.04$ & $\pm0.04$ & -- \\
            [0.3em]
        \hline
        \hline
        \multicolumn{6}{|c|}{\textbf{LVDT response (V/mmV) and data-driven correction factor}} \\
        \hline
        $s_{\rm osc}$& \multicolumn{5}{c|}{$0.01204^{+0.00004}_{-0.00006}$} \\
        $s_{\rm sim}$ & \multicolumn{5}{c|}{$0.01189^{+0.00009}_{-0.00004}$} \\
        $s_{\rm daq}$& \multicolumn{5}{c|}{$0.88114^{+0.00121}_{-0.00121}$}\\
        \hline
        \hline
        $C_{\rm ds}$ & \multicolumn{5}{c|}{$73.18^{+0.26}_{-0.38}$}\\
        \hline
    \end{tabular}
\end{table}
\section{VC measurements with spring-balance method}
\label{sec:VC_measurement}

To measure the force generated by the VC actuator of the ETpathfinder Type-A assembly the developed spring-balance system, described in section~\ref{sec:exp-setup}, will be used. A low frequency (0.1~Hz) sinusoidal voltage signal is generated through the DAC module, and sent to the VC driver on the electronics circuit to convert this into a low frequency current signal. The secondary coils, which are mounted on the suspension frame, start to oscillate vertically due to the generated Lorentz force, while the primary coil (and permanent magnet) remains fixed at a certain position. The springs will transfer this vertical oscillation to the balance, which measures a change in weight ($\Delta m$). With the aforementioned formula, $\mathbf{F} = k\Delta m\mathbf{g}$, the force generated by the VC actuator can then be extracted. The choice of a low frequency actuation signal ensures that the balance weight measurement remains accurate. When the frequency is too high, the sinusoid movement will not be properly recorded by the balance, and a wrong force value will be obtained. To extract the amplitude $\Delta m$, the balance data is fitted with a sine wave function, similar to that used for the LVDT response measurement (section~\ref{sec:LVDT}). Finally, to obtain the VC force as a function of primary coil position this procedure is repeated at each measurement point, after moving the primary coil with the vertical translation stage.

\begin{figure}[ht]
    \centering
    \includegraphics[width=0.55\linewidth]{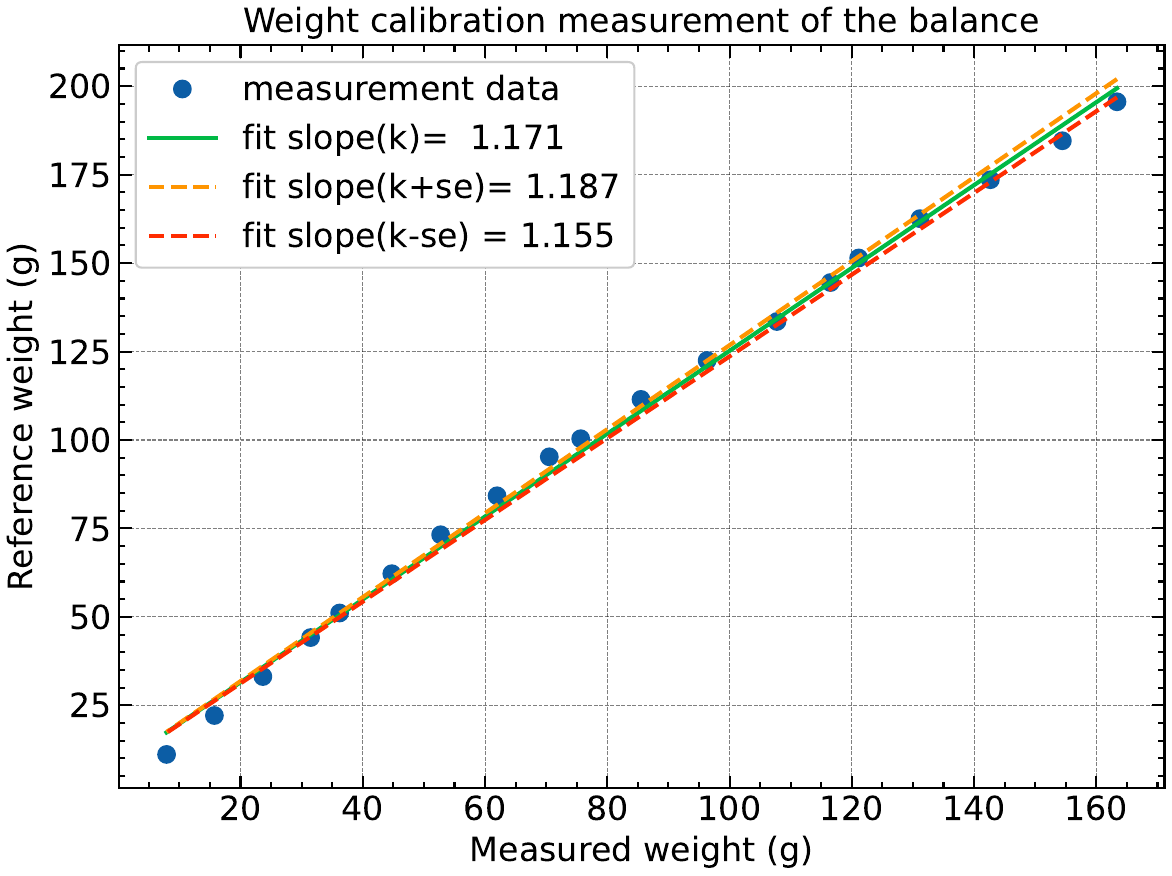}
    \caption{The weight calibration measurement, and linear regression to obtain the $k$-factor. The x-axis denotes the measured weight calculated from the weight difference, and the y-axis denotes the actual weight of the reference masses.}
    \label{fig:calibration_k_factor}
\end{figure}

Due to the springs and suspension frame the balance readout does not represent the actual weight change. This is mitigated by the correction factor $k$ in the formula. Instead of solely relying on the spring-constant value specified by the manufacturer, we determine this factor experimentally to take into account all possible mechanical effects~\cite{pengbo-thesis}. This is done by using a set of reference masses, and first to cross-check their actual weights using the balance without the suspension system. The second step involves combining those reference masses to create different mass blocks, and placing those on the suspension platform with the secondary coils installed. First, the weight is measured without the reference masses added (i.e. weight of secondary coils and suspension platform), then with the references masses. By subtracting the latter and the former, the observed weight difference is obtained, reflecting how much weight is transmitted to the balance via the suspension frame. Comparing this with the actual weights for different combinations of reference masses allows the calculation of the correction factor $k$. The results are shown in figure~\ref{fig:calibration_k_factor}, which plots the reference weight as a function of measured weight in a range of $8-163$~g. With linear regression and standard error calculation the correction factor $k$ is then determined to be $k = 1.171 \pm 0.016$. 

Four measurements of the Type-A actuator were performed, each over a range in $z$ of 25~mm around the primary coil central position with a step size of 1~mm. A 9.9~V signal with $f = 0.1$~Hz is sent to the VC driver circuit, and converted into a current signal with an amplitude of 59.6~mA. At each $z$-value the balance is read out for a certain time (few minutes) to obtain a measurement of the actuation oscillations. This data is subsequently fitted with a sine wave function to extract $\Delta m$. The normalised actuation force (N/A) is then obtained by multiplying this value with the $k$-factor, the gravitational acceleration $\mathbf{g}$, and dividing it with the actuator current.

\begin{figure}[ht]
    \centering
    \includegraphics[width=0.65\linewidth]{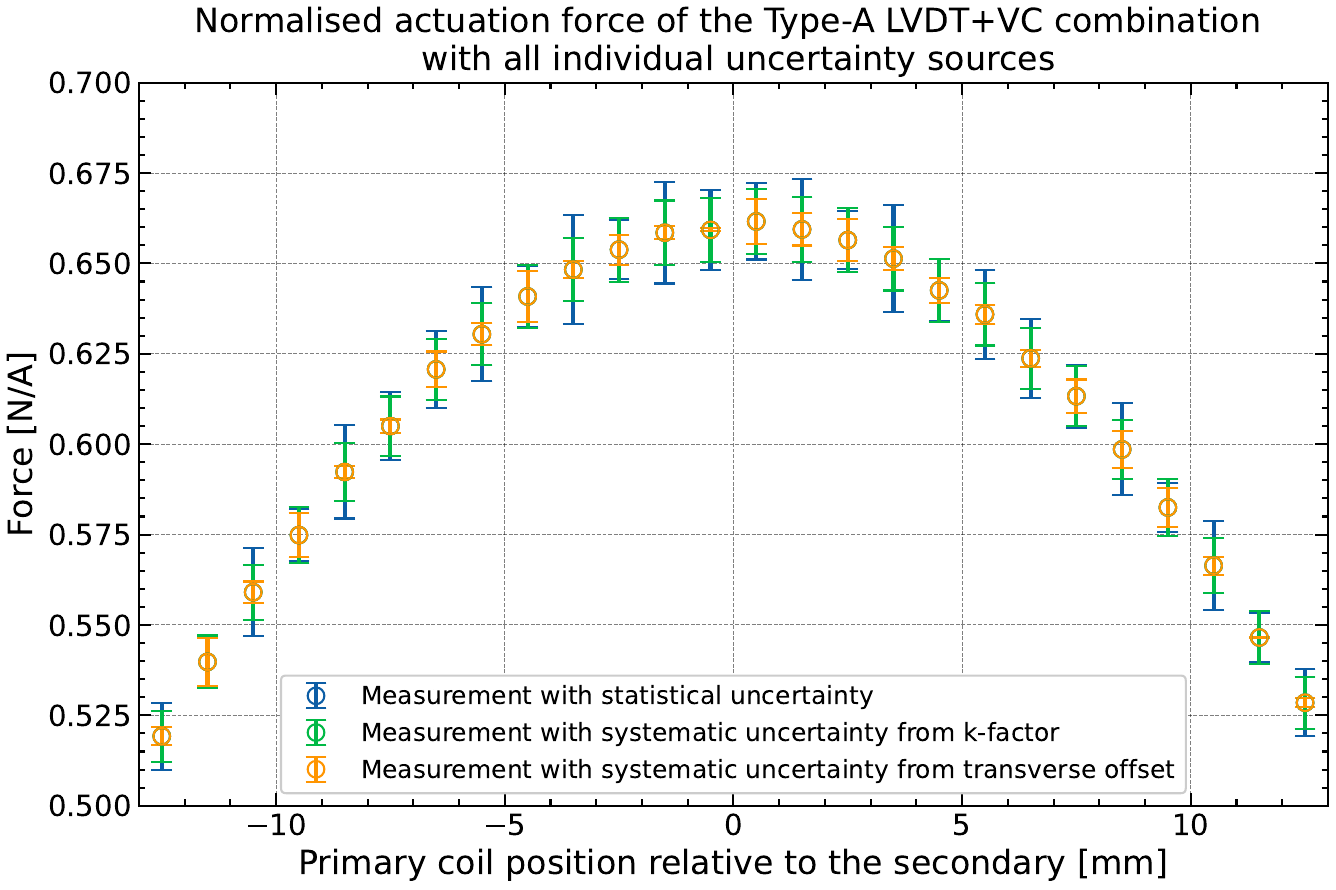}
    \caption{The measured normalised actuation force (N/A) of the Type-A LVDT+VC when driven by a 0.1~Hz current signal of 59.6~mA, with all statistical and systematic errors shown independently. The statistical uncertainty leads to the largest contribution in most data points, while the transverse offset uncertainty strongly fluctuates with primary coil position.}
    \label{fig:vc-measurement-err-comparison}
\end{figure}

The statistical and systematic uncertainties are evaluated using the same methods as outlined in section~\ref{sec:systematics}. For the VC force measurement, two systematic sources are considered: the transverse offset, and the $k$-factor. The contribution of the former is obtained exactly the same as before, while the latter is obtained by varying the $k$-value with its standard error. The individual contributions of the statistical and systematic uncertainties are shown in figure~\ref{fig:vc-measurement-err-comparison}. In contrast to the LVDT measurement, the statistical uncertainties (blue) on the VC force measurement are significant, and range from 0.33~\% to 3.37~\%. This can be attributed to several factors as the spring-balance method is inherently sensitive to environmental perturbations such as mechanical vibrations, air currents, and slow temperature drifts, which can induce fluctuations in the recorded signal. In addition, the finite mechanical response time, damping of the balance, and changes in alignment of the coils can introduce amplitude fluctuations across repeated measurements. These effects can also lead to correlated contributions to the statistical errors on the data points. The uncertainty associated with the correction factor $k$ (green) exhibits a constant deviation of 1.35\%, while the transverse offset uncertainty (orange) ranges from 0.02\% to 14.52\%, depending on the position of the primary coil. 

\begin{figure}[ht]
    \centering
    \includegraphics[width=0.65\linewidth]{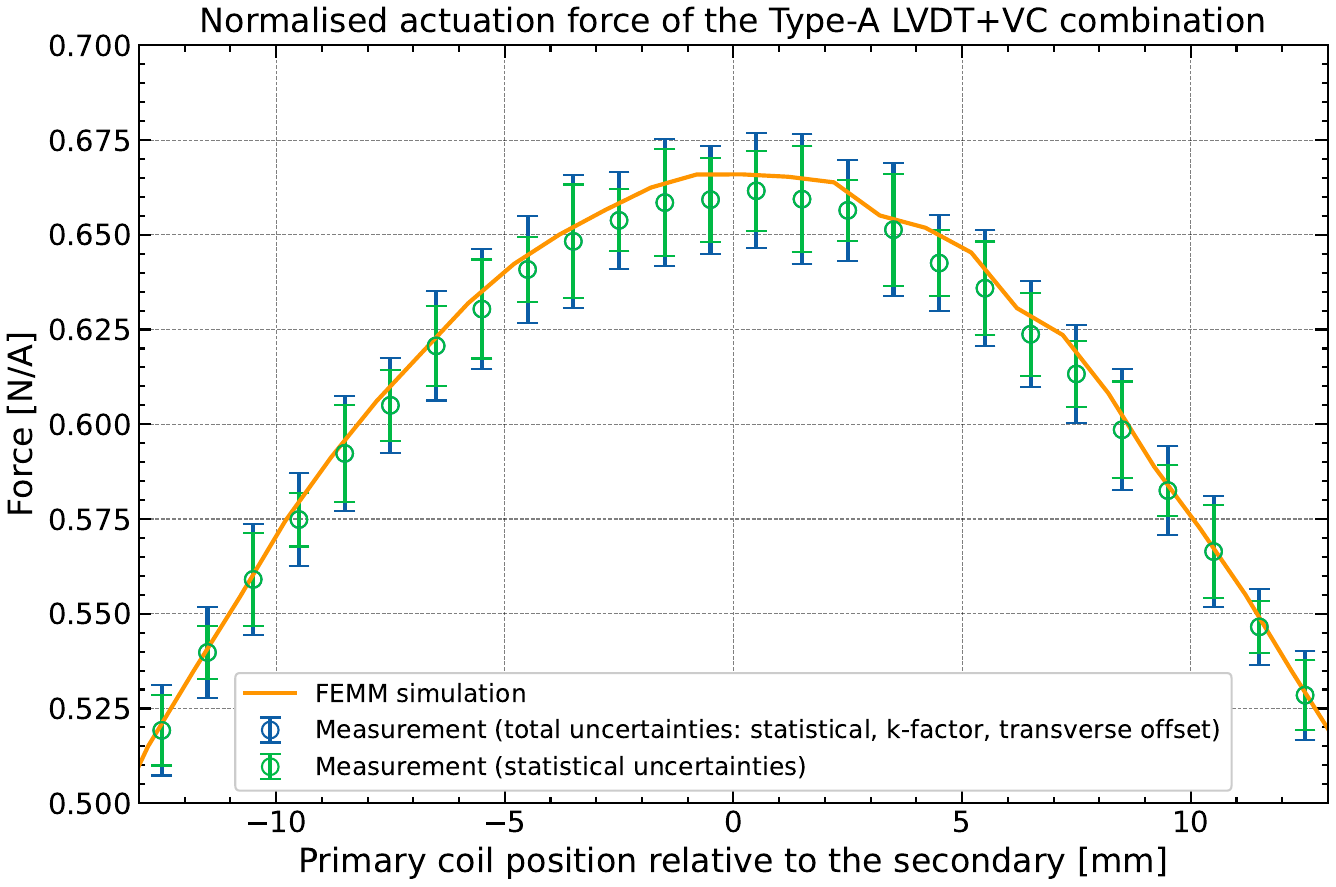}
    \caption{Comparison of the Type-A LVDT+VC normalised actuation force (N/A) between simulation (solid line) and measurement (data points), including both the statistical uncertainty and the total uncertainty (statistical and systematic combined). During the measurement the actuator is driven by a 0.1~Hz current signal of 59.6~mA. The FEMM simulation (with magnet $H_{c} = 969969~\mathrm{A/m}$) agrees very well with the measurement over a 25~mm range of primary coil positions.}
    \label{fig:vc-measurement-simulation}
\end{figure}

The final results are presented in figure~\ref{fig:vc-measurement-simulation}, where the measurement of the normalised VC actuation force (data points) is compared with predictions from the FEMM simulation (solid line), as described in section~\ref{sec:working_principle}. The data points are plotted with the statistical errors only (green), and with the total uncertainties (blue) that range from 1.61~\% to 14.62~\%. In these simulations, the N40 magnet material was directly taken from the built-in FEMM library, where the coercivity is defined by a nominal reference value ($H_{c} = 969969~\mathrm{A/m}$). The simulated forces agree very well within the uncertainties of the measurement at all primary coil positions, which demonstrates that the developed spring-balance method is able to correctly extract an actuation force over a wide range in $z$. At the central position of the primary coil, where the maximum force is generated, the measurement reaches $0.661 \pm 0.015\ (2.3\%) ~\mathrm{N/A}$, while the simulation predicts $0.665~\mathrm{N/A}$. This corresponds to a discrepancy of only 0.6\%, well within the total uncertainty. In addition, within $\pm 5$~mm the generated force remains stable, only decreasing by 4\%.
These measurements confirm that the ETpathfinder Type-A LVDT+VC combination can be used as a proper actuator, delivering a stable VC force that is maximal when the primary coil is in the central position.

\section{Conclusions}
\label{sec:conclusions}
We have developed and validated a dedicated experimental and simulation framework for the detailed characterisation of a combined LVDT position sensor and VC actuator designed for seismic isolation systems in gravitational wave detectors. Using a precision measurement setup together with finite-element modelling (based on FEMM), the ETpathfinder Type-A LVDT+VC assembly was systematically investigated in both sensing and actuation modes, and the experimental results were quantitatively compared with model predictions.

For the LVDT sensing mode, a response of $s_{\mathrm{osc}} = 0.01204^{+0.00004}_{-0.00006} \, \mathrm{V/mmV}$ was measured at the pre-amplification stage, in agreement with the FEMM prediction of $s_{\mathrm{sim}} = 0.01189^{+0.00009}_{-0.00004} \, \mathrm{V/mmV},$ corresponding to a difference of only 1.3\%. The precision of the measurement is within 0.5\%, while the uncertainty on simulations remains within 0.8\%. After including a data-driven correction factor to account for the analogue amplification and digital signal processing chain, the post-amplification response was consistently reproduced by the corrected simulation model. The LVDT sensor exhibits an excellent linearity over a $\pm$5~mm displacement range, with deviations below 1\%, confirming its suitability for precision low-frequency suspension control. The excitation voltage has the largest systematic uncertainty (up to 0.25\%) during measurements, while the mesh size configuration dominates the systematic uncertainty (up to 0.54\%) in simulations.

For the VC actuation mode, the measured normalised force at the central position reaches $0.661 \pm 0.015\ (2.3\%) \ \mathrm{N/A}$, in very good agreement with the FEMM prediction of 0.665 N/A when the magnet coercivity is determined from the built-in material library. This corresponds to a discrepancy of 0.6\%, well within the combined statistical and systematic uncertainties. The actuation force remains stable within $\pm$5~mm of the central position, decreasing only 4\%, thereby providing reliable control across the relevant dynamic range. Despite the environmental sensitivity due to the spring-balance method, the overall agreement validates both the electromagnetic modelling and the force measurement methodology.

Beyond the validation of the ETpathfinder Type-A design, an important outcome of this work is the establishment of a versatile and well-controlled experimental platform that can now be used for future research projects. The setup enables systematic testing, optimisation, and direct simulation validation of novel LVDT sensor geometries and VC actuator prototypes, including studies of linearity, dynamic range, environmental coupling, and electronic readout performance. In combination with the validated FEMM simulation framework, it provides a powerful tool for novel design studies and quantitative performance assessment.

These results demonstrate that the characterised LVDT+VC combination performs in agreement with simulation predictions, and can be used for low-frequency local control in advanced seismic isolation systems. The developed experimental framework directly supports ongoing optimisation efforts for next-generation gravitational wave observatories such as the Einstein Telescope, where precise modelling and minimisation of sensor and actuator limitations can contribute to improved low-frequency strain sensitivity.

\acknowledgments

The simulation code used to model the Type-A LVDT+VC assembly is available here: \url{https://github.com/beginner117/LVDT-VC-modelling-toolkit/tree/simulation_paper}. 
The authors would like to thank the researchers at Nikhef and the ETpathfinder collaboration, especially Fred Schimmel, Alessandro Bertolini, Bas Swinkels, Mathĳs Baars, Gino Hoft, and Stefan Hild for the valuable discussions, feedback on the experimental setup design, and the Type-A design drawings.

The ETpathfinder project in Maastricht is funded by Interreg Vlaanderen-Nederland, the province of Dutch Limburg, the province of Antwerp, the Flemish Government, the province of North Brabant, the Smart Hub Flemish Brabant, the Dutch Ministry of Economic Affairs, the Dutch Ministry of Education, Culture and Science, and by own funding of the involved partners.
In addition the ETpathfinder team acknowledges support from the European Research Council (ERC), the Dutch Research Council (NWO), the Research Foundation Flanders (FWO), the German Research Foundation (DFG), Spanish MICINN, the CERCA program of the Generalitat de Catalunya and the Dutch National Growth Fund (NGF).



\bibliographystyle{JHEP}
\bibliography{references}

@article{tariq2002linear,
  title={{The linear variable differential transformer (LVDT) position sensor for gravitational wave interferometer low-frequency controls}},
  author={Tariq, Hareem and Takamori, Akiteru and Vetrano, Flavio and Wang, Chenyang and Bertolini, Alessandro and Calamai, Giovanni and DeSalvo, Riccardo and Gennai, Alberto and Holloway, Lee and Losurdo, Giovanni and others},
  journal={Nucl. Instrum. Methods Phys. Res., Sect. A},
  volume={489},
  number={1-3},
  pages={570-576},
  year={2002},
  publisher={Elsevier}
}

@misc{femm,
  author       = {D. C. Meeker},
  title        = {Finite {E}lement {M}ethod {M}agnetics},
  howpublished = {Version 4.2 (21Apr2019 Build)},
  year         = {2019},
  url          = {http://www.femm.info},
}

@misc{pyfemm,
    title = { {Python interface to Finite Element Method Magnetics (FEMM)} },
    author = {D. C. Meeker},
    howpublished = {pyfemm 0.1.3},
    year = {2021}, 
    url = {https://www.femm.info/wiki/pyFEMM/manual.pdf}
}

@article{VirgoSuperattenuator,
  author = {Ballardin, G. and Bracci, L. and Braccini, S. and Bradaschia, C. and Casciano, C. and Calamai, G. and Cavalieri, R. and Cecchi, R. and Cella, G. and Cuoco, E. and others},
  title = {{Measurement of the VIRGO superattenuator performance for seismic noise suppression}},
  journal = {Review of Scientific Instruments},
  volume = {72},
  number = {9},
  pages = {3644--3652},
  year = {2001},
  doi = {10.1063/1.1392338}
}

@article{SteeringFilter,
    author = {Ballardin, G. and Braccini, S. and Bradaschia, C. and Casciano, C. and Cavalieri, R. and Cecchi, R. and Chickarmane, V. and others},
    title = {{Measurement of the transfer function of the steering filter of the Virgo super attenuator suspension}},
    journal = {Review of Scientific Instruments},
    volume = {72},
    number = {9},
    pages = {3635-3642},
    year = {2001},
    month = {09},
    issn = {0034-6748},
    doi = {10.1063/1.1384426},
    url = {https://doi.org/10.1063/1.1384426},
}

@article{LowNoiseAccelerometer,
  author = {Braccini, S. and Bradaschia, C. and Del Fabbro, R. and Di Virgilio, A. and Ferrante, I. and Fidecaro, F. and Flaminio, R. and Gennai, A. and Giassi, A. and Giazotto, A. and others},
  title = {Low noise wideband accelerometer using an inductive displacement sensor},
  journal = {Review of Scientific Instruments},
  volume = {66},
  number = {1},
  pages = {741--743},
  year = {1995},
  doi = {10.1063/1.1145608}
}

@inproceedings{losurdoActiveControlsInterferometric2000,
  title = {Active Controls in Interferometric Detectors of Gravitational Waves: Inertial Damping of the {{VIRGO}} Superattenuator},
  shorttitle = {Active Controls in Interferometric Detectors of Gravitational Waves},
  booktitle = {Experimental {{Physics}} of {{Gravitational Waves}}},
  author = {Losurdo, Giovanni},
  year = 2000,
  month = oct,
  eprint = {gr-qc/0002006},
  pages = {379--389},
  doi = {10.1142/9789812792846_0014},
  urldate = {2026-05-23},
  archiveprefix = {arXiv},
}

@article{maggioreAngularControlNoise2025,
  title = {Angular Control Noise in {{Advanced Virgo}} and Implications for the {{Einstein Telescope}}},
  author = {Maggiore, Riccardo and Freise, Andreas and Perry, Jonathan W. and {Mow-Lowry}, Conor M. and Ruggi, Paolo and Mantovani, Maddalena and Casanueva Diaz, Julia and Brown, Daniel and Mart{\'i}n, Enzo N. Tapia San and Tacca, Matteo and Bersanetti, Diego},
  year = 2025,
  journal = {Physical Review D},
  volume = {111},
  number = {10},
  pages = {102003},
  publisher = {American Physical Society},
  doi = {10.1103/PhysRevD.111.102003},
}

@phdthesis{kumar-thesis,
  author       = {Kumar Akhil Kukkadapu},
  title        = {{Development of low frequency position sensors \& actuators, for gravitational wave detectors, and their application in seismic controls in ETpathfinder}},
  school       = {University of Antwerp},
  year         = {2026},
  address      = {Antwerp, Belgium},
  doi          = {},
  url          = {},
  note         = {PhD thesis, Faculty of Science},
}

@phdthesis{pengbo-thesis,
  author       = {Pengbo Li},
  title        = {{Characterization and development of novel position sensors and actuators for seismic attenuation systems in gravitational wave detectors}},
  school       = {University of Antwerp},
  year         = {2025},
  address      = {Antwerp, Belgium},
  doi          = {10.63028/10067/2187960151162165141},
  url          = {https://doi.org/10.63028/10067/2187960151162165141},
  note         = {PhD thesis, Faculty of Science},
}

@article{WANG2002563,
title = {{Constant force actuator for gravitational wave detector's seismic attenuation systems (SAS)}},
journal = {Nucl. Instrum. Methods Phys. Res. A},
volume = {489},
number = {1},
pages = {563-569},
year = {2002},
issn = {0168-9002},
doi = {https://doi.org/10.1016/S0168-9002(02)00801-X},
url = {https://www.sciencedirect.com/science/article/pii/S016890020200801X},
author = {Chenyang Wang and Hareem Tariq and Riccardo DeSalvo and Yukiyoshi Iida and Szabolcs Marka and Yuhiko Nishi and Virginio Sannibale and Akiteru Takamori},
}

@article{STOCHINO2009737,
title = {{The Seismic Attenuation System (SAS) for the Advanced LIGO gravitational wave interferometric detectors}},
journal = {Nucl. Instrum. Methods Phys. Res. A},
volume = {598},
number = {3},
pages = {737-753},
year = {2009},
issn = {0168-9002},
doi = {https://doi.org/10.1016/j.nima.2008.10.023},
url = {https://www.sciencedirect.com/science/article/pii/S0168900208015064},
author = {Alberto Stochino and Benjamin Abbot and Yoichi Aso and Mark Barton and Alessandro Bertolini and Valerio Boschi and Dennis Coyne and Riccardo DeSalvo and Carlo Galli and Yumei Huang and Alex Ivanov and Szabolcs Marka and David Ottaway and Virginio Sannibale and Chiara Vanni and Hiroaki Yamamoto and Sanichiro Yoshida},
}

@article{Acernese_2014,
   title={{Advanced Virgo: a second-generation interferometric gravitational wave detector}},
   volume={32},
   ISSN={1361-6382},
   url={http://dx.doi.org/10.1088/0264-9381/32/2/024001},
   DOI={10.1088/0264-9381/32/2/024001},
   number={2},
   journal={Classical and Quantum Gravity},
   publisher={IOP Publishing},
   author={Acernese, F and Agathos, M and Agatsuma, K and Aisa, D and Allemandou, N and Allocca, A and Amarni, J and Astone, P },
   year={2014},
   month=dec, pages={024001} }

@article{10.1063/1.4866659,
    author = {Beker, M. G. and Bertolini, A. and van den Brand, J. F. J. and Bulten, H. J. and Hennes, E. and Rabeling, D. S.},
    title = {{State observers and Kalman filtering for high performance vibration isolation systems}},
    journal = {Review of Scientific Instruments},
    volume = {85},
    number = {3},
    pages = {034501},
    year = {2014},
    month = {03},
    issn = {0034-6748},
    doi = {10.1063/1.4866659},
    url = {https://doi.org/10.1063/1.4866659},
}

@article{Heijningen:2019jmd,
    author = "van Heijningen, J. V and Bertolini, A. and Hennes, E. and Beker, M. G. and Doets, M. and Bulten, H. J. and Agatsuma, K. and Sekiguchi, T. and Van Den Brand, J. F J.",
    title = "{{A multistage vibration isolation system for Advanced Virgo suspended optical benches}}",
    doi = "10.1088/1361-6382/ab075e",
    journal = "Class. Quant. Grav.",
    volume = "36",
    number = "7",
    pages = "075007",
    year = "2019"
}

@article{Kirchhoff:2020vhb,
    author = "Kirchhoff, R. and others",
    title = "{{Local active isolation of the AEI-SAS for the AEI 10 m prototype facility}}",
    doi = "10.1088/1361-6382/ab857e",
    journal = "Class. Quant. Grav.",
    volume = "37",
    number = "11",
    pages = "115004",
    year = "2020"
}

@article{Akutsu:2021auw,
    author = "Akutsu, T. and others",
    title = "{{Vibration isolation systems for the beam splitter and signal recycling mirrors of the KAGRA gravitational wave detector}}",
    doi = "10.1088/1361-6382/abd922",
    journal = "Class. Quant. Grav.",
    volume = "38",
    number = "6",
    pages = "065011",
    year = "2021"
}

@article{Punturo:2010zz,
    author = "Punturo, M. and others",
    editor = "Ricci, Fulvio",
    title = "{{The Einstein Telescope: A third-generation gravitational wave observatory}}",
    doi = "10.1088/0264-9381/27/19/194002",
    journal = "Class. Quant. Grav.",
    volume = "27",
    pages = "194002",
    year = "2010"
}

@techreport{et_design_report_update2020,
  title        = {{Design Report Update 2020 for the Einstein Telescope}},
  author       = {{ET Steering Committee Editorial Team}},
  institution  = {Einstein Telescope Collaboration},
  year         = {2020},
  type         = {Technical Design Report},
  url          = {https://www.einsteintelescope-emr.eu/wp-content/uploads/2024/05/ET-0007B-20_ETDesignReportUpdate2020.pdf},
  note         = {Published September 2020},
}

@book{ETpathfinderTDR,
  author       = "{The ETpathfinder Team}",
  title        = "{{ETpathfinder Design Report}}",
  publisher = "ET-0011A-20",
  year         = "2020",
  url={https://etpathfinder.eu/wp-content/uploads/2020/03/
ETpathfinder-Design-Report.pdf},
}

@article{Utina:2022qqb,
    author = "Utina, A. and others",
    title = "{{ETpathfinder: a cryogenic testbed for interferometric gravitational-wave detectors}}",
    eprint = "2206.04905",
    archivePrefix = "arXiv",
    primaryClass = "astro-ph.IM",
    doi = "10.1088/1361-6382/ac8fdb",
    journal = "Class. Quant. Grav.",
    volume = "39",
    number = "21",
    pages = "215008",
    year = "2022"
}

@book{Jackson:1998nia,
    author = "Jackson, John David",
    title = "{Classical Electrodynamics}",
    isbn = "978-0-471-30932-1",
    publisher = "Wiley",
    year = "1998"
}

@book{Griffiths:1492149,
      author        = "Griffiths, David J",
      title         = "{Introduction to electrodynamics}",
      publisher     = "Pearson",
      year          = "2013",
}

@book{nyce:2016,
author = {Nyce, David S.},
address = {Hoboken},
edition = {2nd},
isbn = {9781119069362},
publisher = {Wiley},
title = {Position Sensors : Theory and Application.},
year = {2016},
}

@INPROCEEDINGS{virgo_daq_lapp,
  author={Acernese, F. and Amico, P. and Alshourbagy, M. and Antonucci, F. and Aoudia, S. and Astone, P. and Avino, S. and Babusci, D. and Ballardin, G. and Barone, F. and Barsotti, L. and Barsuglia, M. and Bauer, Th. S. and Beauville, F. and Bigotta, S. and Birindelli, S. and Bizouard, M.A. and Boccara, C. and Bondu, F. and Bosi, L. and Bradaschia, C. and Braccini, S. and van den Brand, F. J. and Brillet, A. and Brisson, V. and Buskulic, D. and Calloni, E. and Campagna, E. and Carbognani, F. and Cavalier, F. and Cavalieri, R. and Cella, G. and Cesarini, E. and Chassande-Mottin, E. and Christensen, N. and Corda, C. and Corsi, A. and Cottone, F. and Clapson, A.-C. and Cleva, F. and Coulon, J.-P. and Cuoco, E. and Dari, A. and Dattilo, V. and Davier, M. and del Prete, M. and De Rosa, R. and Di Fiore, L. and Di Virgilio, A. and Dujardin, B. and Eleuteri, A. and Evans, M. and Ferrante, I. and Fidecaro, F. and Fiori, I. and Flaminio, R. and Fournier, J.-D. and Frasca, S. and Frasconi, F. and Gammaitoni, L. and Garufi, F. and Genin, E. and Gennai, A. and Giazotto, A. and Giordano, G. and Giordano, L. and Gouaty, R. and Grosjean, D. and Guidi, G. and Hamdani, S. and Hebri, S. and Heitmann, H. and Hello, P. and Huet, D. and Karkar, S. and Kreckelbergh, S. and La Penna, P. and Lavel, M. and Leroy, N. and Letendre, N. and Lopez, B. and Lorenzini and Loriette, V. and Losurdo, G. and Mackowski, J.-M. and Majorana, E. and Man, C. N. and Mantovani, M. and Marchesoni, F. and Marion, F. and Marque, J. and Martelli, F. and Masserot, A. and Mazzoni, M. and Milano, L. and Menzinger, F. and Moins, C. and Moreau, J. and Morgado, N. and Mours, B. and Nocera, F. and Palomba, C. and Paoletti, F. and Pardi, S. and Pasqualetti, A. and Passaquieti, R. and Passuello, D. and Piergiovanni, F. and Pinard, L. and Poggiani, R. and Punturo, M. and Puppo, P. and van der Putten, S. and Qipiani, K. and Rapagnani, P. and Reita, V. and Remillieux, A. and Ricci, F. and Ricciardi, I. and Ruggi, P. and Russo, G. and Solimeno, S. and Spallicci, A. and Tarallo, M. and Tonelli, M. and Toncelli, A. and Tournefier, E. and Travasso, F. and Tremola, C. and Vajente, G. and Verkindt, D. and Vetrano, F. and Vicere, A. and Vinet, J.-Y. and Vocca, H. and Yvert, M.},
  booktitle={2007 15th IEEE-NPSS Real-Time Conference}, 
  title={Data Acquisition System of the Virgo Gravitational Waves Interferometric Detector}, 
  year={2007},
  volume={},
  number={},
  pages={1-8},
  doi={10.1109/RTC.2007.4382842}}

@misc{man-tektronix_dpo4054,
  title        = {{MSO4000 and DPO4000 Series} Digital Phosphor Oscilloscopes Manual},
  howpublished = {\url{https://www.tek.com/en/oscilloscope/dpo4054-manual/mso4000-and-dpo4000-series-0/}},
  note         = {Tektronix user manual for MSO4000 and DPO4000 Series oscilloscopes (accessed 2026-01-19)},
  organization = {Tektronix, Inc.},
  year         = {2010},
}

@misc{physikinstrumente_website,
  title        = {{Physik Instrumente (PI) – Solutions for Precision Motion and Positioning}},
  howpublished = {\url{https://www.physikinstrumente.com/en/}},
  note         = {Accessed: 2026-01-19},
  organization = {Physik Instrumente (PI) SE \& Co. KG},
  year         = {2026}
}

@misc{pipython_2.11.0.6,
  title        = {{PIPython} v2.11.0.6},
  author       = {{Physik Instrumente (PI) SE \& Co. KG}},
  year         = {2025},
  howpublished = {\url{https://pypi.org/project/PIPython/2.11.0.6/}},
  note         = {Python library to access PI devices and process GCS data (accessed 2026-01-19)},
}

@misc{keyence_lk-g5000,
  title        = {{LK-G5000} Series Laser Displacement Sensor},
  howpublished = {\url{https://www.keyence.eu/products/measure/laser-1d/lk-g5000/}},
  note         = {Accessed: 2026-01-19},
  organization = {KEYENCE International Belgium},
  year         = {2026},
}

@misc{actuonix_L12-50-50-12-I,
  title        = {{Actuonix L12-50-50-12-I} Micro Linear Actuator with Internal Controller},
  howpublished = {\url{https://www.actuonix.com/l12-50-50-12-i}},
  note         = {Accessed: 2026-01-19},
  organization = {Actuonix Motion Devices Inc},
  year         = {2026},
}

@misc{ohaus_px224_balance,
  title        = {{OHAUS PX224} Pioneer™ Analytical Electronic Balance},
  howpublished = {\url{https://pl.ohaus.com/en-ap/products/balances-scales/analytical-balances/pioneer-analytical-(1)/electronic-balance-px224/}},
  note         = {Accessed: 2026-01-19},
  organization = {OHAUS Europe GmbH},
  year         = {2026},
}

@misc{pyserial,
  title        = {{pySerial}: Python Serial Port Extension},
  author       = {Liechti, Chris},
  howpublished = {\url{https://pyserial.readthedocs.io/}},
  note         = {Open-source Python package for serial communication (accessed 2026-01-19)},
  year         = {2026}
}

\end{document}